\documentclass[twocolumn,astrosymb,tighten,trackchanges]{aastex7}

\usepackage{xcolor}
\usepackage{soul}

\begin{document}

\title{A Parsec-Scale Transient High-Frequency-Peaked Radio Source in the Very Large Array Sky Survey Brightening and Fading over a Decadal Timescale}

\correspondingauthor{Yuyang Chen}
\email{yu\_yang.chen@cfa.harvard.edu}

\author[0000-0002-8804-3501]{Yuyang Chen}
\affiliation{David A. Dunlap Department of Astronomy \& Astrophysics, University of Toronto, Toronto, ON M5S 3H4, Canada}
\affiliation{Dunlap Institute for Astronomy \& Astrophysics, University of Toronto, Toronto, ON M5S 3H4, Canada}
\affiliation{Center for Astrophysics $\vert$ Harvard \& Smithsonian, 60 Garden St. Cambridge, MA 02138, USA}
\email{yu_yang.chen@cfa.harvard.edu}

\author[0000-0002-3382-9558]{B. M. Gaensler}
\affiliation{Dunlap Institute for Astronomy \& Astrophysics, University of Toronto, Toronto, ON M5S 3H4, Canada}
\affiliation{David A. Dunlap Department of Astronomy \& Astrophysics, University of Toronto, Toronto, ON M5S 3H4, Canada}
\affiliation{Department of Astronomy and Astrophysics, University of California, Santa Cruz, 1156 High Street, Santa Cruz, CA 95064, USA}
\email{gaensler@ucsc.edu}

\author[0000-0001-7081-0082]{Maria R. Drout}
\affiliation{David A. Dunlap Department of Astronomy \& Astrophysics, University of Toronto, Toronto, ON M5S 3H4, Canada}
\email{maria.drout@utoronto.ca}

\author[0000-0001-9584-2531]{Dillon Z. Dong}
\affiliation{National Radio Astronomy Observatory, P.O. Box O, Socorro, NM 87801, USA}
\email{dongdillon27@gmail.com}

\author[0000-0003-3168-5922]{Emmanuel Momjian}
\affiliation{National Radio Astronomy Observatory, P.O. Box O, Socorro, NM 87801, USA}
\email{emomjian@nrao.edu}

\author[0000-0001-6812-7938]{Tracy Clarke}
\affiliation{U.S. Naval Research Laboratory, 4555 Overlook Ave. SW, Washington, DC 20375, USA}
\email{tracy.e.clarke2.civ@us.navy.mil}

\author[0000-0002-5187-7107]{Wendy Peters}
\affiliation{U.S. Naval Research Laboratory, 4555 Overlook Ave. SW, Washington, DC 20375, USA}
\email{wendy.m.peters8.civ@us.navy.mil}

\author[0000-0003-3272-9237]{Emil Polisensky}
\affiliation{U.S. Naval Research Laboratory, 4555 Overlook Ave. SW, Washington, DC 20375, USA}
\email{emil.j.polisensky.civ@us.navy.mil}

\author[0000-0002-7098-7512]{Ariel D. Amaral}
\affiliation{David A. Dunlap Department of Astronomy \& Astrophysics, University of Toronto, Toronto, ON M5S 3H4, Canada}
\affiliation{Dunlap Institute for Astronomy \& Astrophysics, University of Toronto, Toronto, ON M5S 3H4, Canada}
\email{ariel.amaral@mail.utoronto.ca}

\author[0000-0002-4449-9152]{Katie Auchettl}
\affiliation{School of Physics, The University of Melbourne, VIC 3010, Australia}
\email{katie.auchettl@unimelb.edu.au}

\author[0000-0002-6960-6911]{Ylva G\"{o}tberg}
\affiliation{Institute of Science and Technology Austria (ISTA), Am Campus 1, 3400 Klosterneuburg, Austria}
\email{ylva.gotberg@ista.ac.at}

\author[0000-0002-7329-3209]{Kovi Rose}
\affiliation{Sydney Institute for Astronomy, School of Physics, University of Sydney, NSW, 2006, Australia}
\affiliation{CSIRO Space and Astronomy, PO Box 76, Epping, NSW, 1710, Australia}
\email{kovi.rose@sydney.edu.au}

\author[0000-0002-9132-6561]{Peter Senchyna}
\affiliation{The Observatories of the Carnegie Institution for Science, 813 Santa Barbara Street, Pasadena, CA 91101, USA}
\email{psenchyna@carnegiescience.edu}

\begin{abstract}

A subset of active galactic nuclei (AGN) manifests as compact radio sources with peaked radio spectra, commonly referred to as peaked-spectrum (PS) sources. Generally, PS sources are interpreted as young radio jets, but the early evolution of the jet is rarely observed and not fully understood. Here, we present VT\,J182357-234511 (VT\,J1823-2345), a radio transient in the Very Large Array Sky Survey (VLASS) that resembles a PS source. Notable properties of VT\,J1823-2345 include: (i) brightening by a factor of 3 between 1996 and 2021 at 1.4\,GHz, (ii) peaked spectra with high frequency peaks at $\nu_{\mathrm{p,obs}}\gtrsim6\,\mathrm{GHz}$,
(iii) a peaked 3\,GHz light curve that has been fading since 2023, (iv) a resolved Gaussian (major-axis) size of 0.4\,mas in 2024, and (v) a linear polarization fraction $p_{\mathrm{L}}\simeq0.73\%$ at 8.8\,GHz with an extreme Faraday rotation measure $\mathrm{RM_{\mathrm{obs}}} \sim +12,000\,\mathrm{rad}\,\mathrm{m}^{-2}$. On the sky, VT\,J1823-2345 coincides with a supernova remnant and an F-type star, but no background galaxy is seen. Based on the multi-wavelength properties, we disfavor Galactic transients, explosive extragalactic transients, and blazar variability as plausible explanations. We argue that VT\,J1823-2345 is consistent with a ``short-lived'' parsec-scale PS source 
with a decades-old jet that does not grow into a large-scale radio galaxy. From comparing the full VLASS light curves, we speculate that VT\,J1823-2345 is similar to many VLASS nuclear transients reported in the literature, which may represent an emerging class of transient radio jets, perhaps related to episodic AGN activity or jetted tidal disruption events.
\end{abstract}

\section{Introduction} \label{sec:intro}

Active galactic nuclei (AGN) are among the most luminous and abundant objects on the radio sky. A small subset of compact, radio-loud AGN exhibit concave or peaked radio spectra over meter to centimeter wavelengths ($\sim$\,$0.1$--$10$\,GHz), commonly referred to as peaked-spectrum (PS) sources \citep[][]{ODea1998,ODeaSaikia2021}. PS sources are further divided (empirically) into subcategories based on the peak frequency $\nu_{\mathrm{p}}$. They are often called megahertz-peaked spectrum (MPS) sources if $\nu_{\mathrm{p}}\lesssim 1\,\mathrm{GHz}$, gigahertz-peaked spectrum (GPS) sources if $\nu_{\mathrm{p}}\sim$\,0.5--5$\,\mathrm{GHz}$, and high-frequency peaked (HFP) sources if $\nu_{\mathrm{p}}\gtrsim 5\,\mathrm{GHz}$. PS sources were first discovered several decades ago \citep[e.g., PKS\,1934-638;][]{Kellermann1966}, and over the past decade, thousands of such sources have been identified through various wide-field radio sky surveys \citep[e.g.,][]{Callingham2017,Ross2021,Slob2022,Ballieux2024,Kerrison2025,Majorova2025}.

The spectral turnover in PS sources is generally attributed to absorption, either synchrotron self-absorption (SSA) or free-free absorption (FFA), in two commonly proposed scenarios. SSA is expected to be prominent in the ``youth scenario,'' which suggests that PS sources are young, compact AGN jets \citep[][]{Fanti1995,Snellen2000,An&Baan2012}. In this scenario, the linear size of the jet increases as the jet expands over time, and the peak frequency decreases as the emission gradually becomes optically thin. The youth scenario is supported by very-high angular-resolution very long baseline interferometry (VLBI) observations of PS sources that commonly show symmetric jet structures or lobes at compact (parsec to kiloparsec) physical scales \citep[e.g.,][]{Wilkinson1994,Stanghellini1997,Orienti2006}. Radio measurements also reveal correlations between the peak frequency, linear size, and radio power of PS sources, explainable by young, expanding jets \citep[][]{Fanti1990,ODeaBaum1997,ODea1998,Snellen2000,KunertBajraszewska2010}. In addition, the inferred kinematic and spectral ages of $\lesssim$\,10$^2$--10$^4$\,yr further support the youth scenario for a number of PS sources \citep[e.g.,][]{Tschager2000,Nagai2006,Collier2018,OrientiDallacasa2020,Ross2023}. Within this picture, HFP sources represent the youngest parsec-scale jets that will evolve into kiloparsec-scale GPS and MPS sources, and some of these sources are thought to be precursors of large-scale Fanaroff–Riley galaxies \citep[][]{FanaroffRiley1974}. Thus, they form an important population for understanding the evolution of radio galaxies.

Alternatively, the spectral turnover may be explained by FFA in the ``frustration scenario,'' which suggests that instead of young jets, PS sources are frustrated jets trapped within dense nuclear environments \citep[][]{vanBreugel1984,ODea1991,Peck1999,Bicknell2018}. This scenario is often proposed when FFA provides a better explanation than SSA, especially for PS sources with extremely steep spectra below the peak frequency \citep[e.g.,][]{Callingham2015,Callingham2017,Keim2019,Mhaskey2019}. Such frustrated jets are informative about nuclear environments and feedback processes within galaxies. 

Currently, the youth scenario is thought to be the correct interpretation for the majority of PS sources \citep[][]{ODeaSaikia2021}. However, current samples suggest that PS sources likely make up more than a few percent of the radio-loud AGN population, which implies an overabundance of young PS sources \citep[][]{ODea1998,ODeaSaikia2021,Slob2022,Ballieux2024}. Specifically, since large-scale radio galaxies have ages\,$\sim$\,$10^7$--$10^8$\,yr and PS sources likely have ages\,$\lesssim$\,$10^2$--$10^5$\,yr, if all PS sources were to evolve into large-scale radio galaxies, PS sources should be orders of magnitude less abundant than large-scale radio galaxies. This expectation contradicts the observationally inferred abundance of PS sources of at least a few percent. 

Although the frustration scenario may account (at least partially) for the observed overabundance, an alternative explanation is that some PS sources are young transient jets that do not grow into large-scale radio galaxies \citep[][]{ODeaBaum1997,ReynoldsBegelman1997,Gugliucci2005,Orienti2010,KunertBajraszewska2010,Readhead2024}. These transient PS sources may be driven by intermittent or episodic AGN activities \citep[e.g., due to instabilities;][]{Czerny2009} or even tidal disruption events \citep[TDEs;][]{Readhead2024}. However, direct evidence favoring the transient explanation has been difficult to obtain. In the past, studies generally relied on age constraints inferred from VLBI for a small subset of PS sources \citep[e.g.,][]{Gugliucci2005,Orienti2010} or indirect comparisons on a population level \citep[e.g.,][]{KunertBajraszewska2010} to argue in favor of the transient explanation.

Arguably, one of the most direct ways of confirming the episodic nature of PS sources is to capture the transient event, i.e., the birth or death of the radio source in the form of extreme brightening or fading. Systematic searches for slow-evolving transients on the radio sky have become viable in the most recent decade with the advent of wide-field (slow) time-domain radio surveys. For example, the Karl G. Jansky Very Large Array \citep[VLA;][]{Perley2011JVLA} has conducted the VLA Sky Survey \citep[VLASS;][]{Lacy2020VLASS} at 3\,GHz between 2017 and 2026, covering the entire sky North of $-40^{\circ}$ in declination. VLASS coverage was split into two halves, with the first half observed for four epochs and the second half observed for three epochs, and the cadence was about 32 months. In the southern hemisphere, the Australian SKA Pathfinder \citep[ASKAP;][]{Johnston2007ASKAP,Hotan2021ASKAP} has been conducting surveys since 2019, such as the Rapid ASKAP Continuum Survey \citep[RACS;][]{McConnell2020RACSI,Hale2021RACSIIcatalog,Duchesne2023RACSmidIV,Duchesne2024RACSmidV,Duchesne2025RACShighVI} and the Variables and Slow Transients survey \citep[VAST;][]{Murphy2013VAST,Murphy2021VAST,deRuiter2026VASTextDR1}. RACS has covered the entire sky south of about $+49^{\circ}$ in declination since 2019 in three frequency bands across $\simeq$\,$0.7$--$1.8$\,GHz over multiple epochs (cadence of a few years). The full VAST survey has been observing select patches of sky at 887.5\,GHz since 2022 on a biweekly or monthly cadence and covers over one quarter of the entire sky.

Over the past several years, a number of studies have used these radio surveys to identify individual or samples of radio nuclear transients associated with AGN or TDEs \citep[e.g.,][]{Nyland2020,Ravi2022,Somalwar2022,Zhang2022,Somalwar2023J0243,Anumarlapudi2024,Chen2024InvertedTransient,Dykaar2024,Somalwar2023TDEI,Somalwar2023TDEII,Birmingham2025,French2025,KunertBajraszewska2025}. Some studies have already begun considering possible connections between certain radio nuclear transients and young PS sources \citep[e.g.,][]{KunertBajraszewska2025}.

In this study, we present the discovery of a VLASS transient, J182357-234511 (VT\,J1823-2345 hereafter), which we discovered serendipitously during a search for inverted-spectra radio transients \citep[][]{Chen2024InvertedTransient}. For this search, we made use of VLASS as well as the VLITE Commensal Sky Survey \citep[VCSS;][]{Clarke2016,Clarke2016VLITE,Polisensky2016,Polisensky2019VDP,Peters2021VCSS,Polisensky2024}, which observed simultaneously with VLASS at 340\,MHz and provided useful spectral information for transient objects. VT\,J1823-2345 was identified as a candidate radio transient possibly associated with an unusual Galactic transient X-ray binary. Following this discovery, we initiated a variety of observing programs to obtain multi-wavelength constraints on its properties. Here, we present these observations and discuss the nature of VT\,J1823-2345, considering both Galactic and extragalactic origins. In particular, we focus on the interpretation that VT\,J1823-2345 is a young PS source.

This paper is structured as follows. In Section~\ref{sec:obs}, we describe the discovery of VT\,J1823-2345 and outline the follow-up and archival observations used in this study. In Section~\ref{sec:analysis}, we present analyses of the observed properties. In Section~\ref{sec:episodicAGN}, we discuss the interpretation that VT\,J1823-2345 is a ``short-lived'' transient PS source, which we consider to be the most plausible scenario. Note that we also discuss a list of alternative, unlikely scenarios, including various Galactic and extragalactic transients, and present arguments disfavoring them in Appendix~\ref{sec:unlikelyscene}. We summarize our results and conclude in Section~\ref{sec:conclusion}. Throughout this study, we adopt a flat Universe with cosmological parameters $H_0 = 67.4\,\mathrm{km}\,\mathrm{s}^{-1}\,\mathrm{Mpc}^{-1}$ and $\Omega_\mathrm{m}=0.315$ \citep[][]{Plank2018VICosmoPar}.

\section{Observations} \label{sec:obs}

In this section, we describe our discovery of VT\,J1823-2345 and follow-up observations obtained in the radio, X-ray, and optical. We also describe a collection of archival and survey data used to expand our spectral and temporal coverage of VT\,J1823-2345.

\subsection{Discovery of VT\,J1823-2345}

\begin{figure*}
    \epsscale{1.0}
    \centering
    \plotone{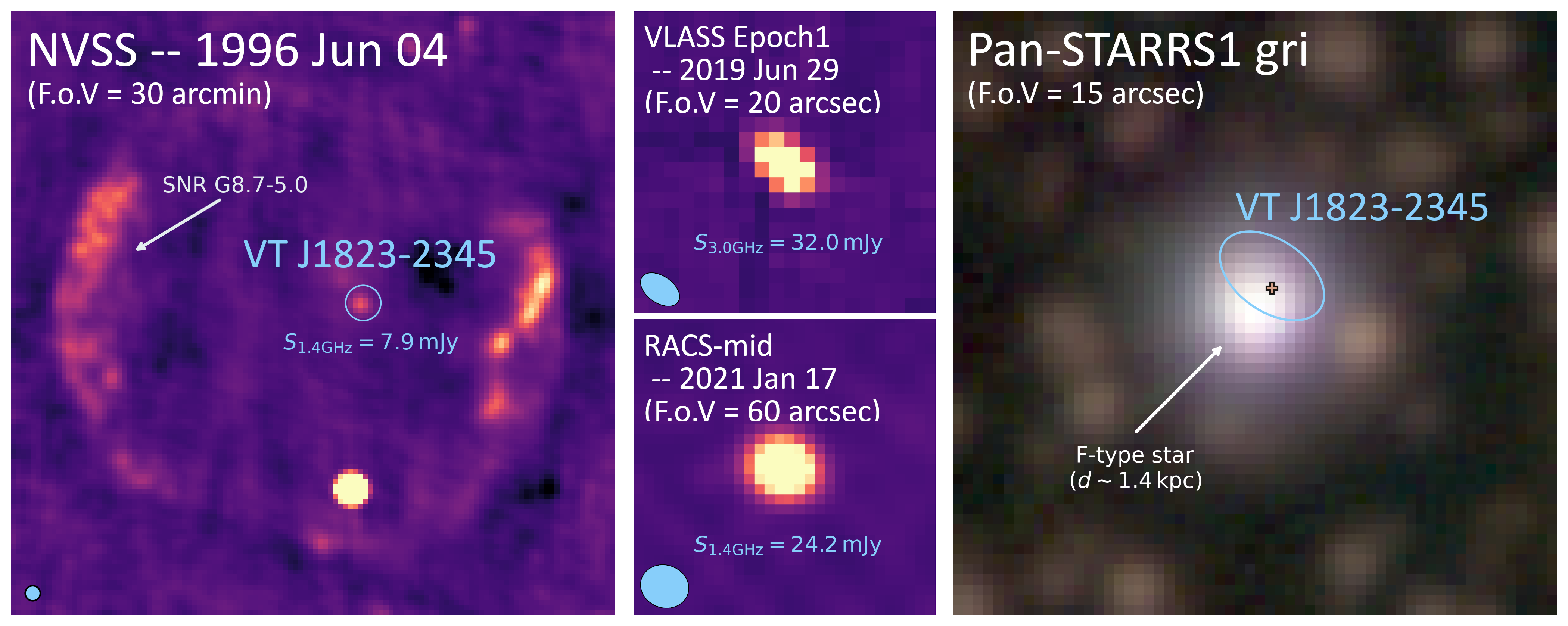}
    \caption{\textit{Left and middle panels}: Radio images from NVSS, VLASS epoch 1, and RACS-mid1 showing VT\,J1823-2345, with observation dates and field of view (F.o.V) shown at the top left. Restoring beams of the radio images are shown at the bottom left while the radio flux density of VT\,J1823-2345 is shown below the source. The NVSS image also shows SNR G8.7-5.0. \textit{Right panel}: Three-color ($g$, $r$, $i$-band) Pan-STARRS1 image showing spatial coincidence between VT\,J1823-2345 and an F-type star. The ellipse is the radio beam from VLASS epoch 1 while the cross marks the position obtained from VLBA observations (accurate to $<$\,1\,mas).}
    \label{fig:radoptimg}
\end{figure*}

We discovered VT\,J1823-2345 during a search for transients with inverted spectra in VLASS and VCSS, two surveys conducted simultaneously by the VLA that began in 2017. The details of the search are described by \citet{Chen2024InvertedTransient}. In summary, we searched for transient candidates that were detected in the first epoch of VLASS (between 2017--2019) but not detected in past radio surveys, such as the NRAO VLA Sky Survey \citep[NVSS;][]{Condon1998NVSS} conducted over 1993--1996 at 1.4\,GHz. By combining the simultaneous coverage of VLASS (3.0\,GHz) and VCSS (340\,MHz), we derived spectral indices $\alpha$ (assuming a power law spectrum of the form $S_\nu \propto \nu^\alpha$, where $S_\nu$ is the flux density) and selected transient candidates with inverted spectra, i.e., $\alpha>0$. The motivation behind this selection is that an inverted spectrum for a transient could be an indicator of interesting properties such as dynamical youth and extreme absorption \citep{Chen2024InvertedTransient}. 

During this search, we discovered VT\,J1823-2345 as a possible inverted-spectrum variable source that had a flux density of $S_{1.4\mathrm{GHz}} = 7.9\,\mathrm{mJy}$ in NVSS in 1996 and $S_{3.0\mathrm{GHz}} = 32.0\,\mathrm{mJy}$ in VLASS epoch 1 in 2019, but it was below the detection threshold ($5\sigma$ above image rms) of the VCSS epoch 1 catalog. Although the detection in NVSS led to its exclusion from the final transient candidate list of \citet{Chen2024InvertedTransient}, its variability was hinted by the flux density difference between NVSS and VLASS. In the first epoch of the RACS mid-band survey \citep[RACS-mid1, taken at 1367.5\,MHz;][]{Duchesne2023RACSmidIV,Duchesne2024RACSmidV}, VT\,J1823-2345 showed a flux density of $S_{1.4\mathrm{GHz}} \approx 24.2\,\mathrm{mJy}$ in 2021, which confirmed that it brightened by a factor of three within a few decades. For radio galaxies or AGN, such brightening may imply a transient source \citep[e.g.,][]{Nyland2020}. With a flux upper limit of about 15\,mJy, the VCSS non-detection implies that $\alpha\gtrsim0.35$ between 340\,MHz and 3.0\,GHz. The radio survey images are shown in Figure~\ref{fig:radoptimg}.

An intriguing aspect of VT\,J1823-2345 is its location on the sky. VT\,J1823-2345 is near the projected center of the supernova remnant (SNR) G8.7-5.0 \citep[][]{Reich1988SNR} and its VLASS epoch 1 position was only 1.16\,arcsec away from an F-type star \citep[Gaia DR3 4089337082410644992;][]{GaiaMission2016,GaiaDR32023} after accounting for proper motion (see Figure~\ref{fig:radoptimg}). Since the astrometric uncertainty of VLASS can be large ($\gtrsim$\,1\,arcsec) at low declinations \citep[][]{Gordon2021VLASS}, an association between VT\,J1823-2345 and the F-type star should be considered. The Gaia DR3 distance of the F-type star is $d\sim1.4\,\mathrm{kpc}$, which would imply a high radio luminosity of $\nu L_{\nu} = 10^{29}\,\mathrm{erg}\,\mathrm{s}^{-1}$ at $\nu=3.0\,\mathrm{GHz}$ in VLASS epoch 1 that is consistent with X-ray binaries \citep[][]{Tetarenko2016}. Therefore, we identified VT\,J1823-2345 as a candidate nearby transient X-ray binary that could potentially be linked to a SNR. Based on this speculation, we initiated a variety of multi-wavelength follow-up observations to characterize the nature of VT\,J1823-2345. 

\subsection{Radio Observations}

\subsubsection{ATCA}

To constrain the radio continuum, we observed VT\,J1823-2345 using the Australia Telescope Compact Array (ATCA) on 2023 January 21 in the 6C configuration under program CX526 (Director's Time). The \textsc{cfb-1m} mode was used for the Compact Array Broadband Backend correlator to record visibilities at 16\,cm and 4\,cm in three 2048\,MHz bands (1\,MHz resolution) centered on 2.1\,GHz, 5.5\,GHz, and 9.0\,GHz. The observation spanned 12\,hr and included scans of the primary calibrator PKS\,1934-638, the nearby complex gain calibrator 1817-254, and a nearby check source NVSS\,J182359-235423. Observing bands were alternated (between 16\,cm and 4\,cm) throughout the observing session to maximize the $(u,v)$-coverage in all bands. 

We processed the ATCA data using \texttt{Miriad} \citep{Sault1995Miriad} following standard reduction procedure. After initial flagging of radio frequency interference, we noticed that some scans on certain baselines exhibited significant phase decorrelation, presumably caused by poor weather conditions or atmospheric turbulence. These scans were further flagged to avoid introducing significant phase errors in calibration and imaging. Calibration solutions were derived from the primary and gain calibrators for all bands (2.1\,GHz, 5.5\,GHz, 9.0\,GHz), and were applied to both our target VT\,J1823-2345 and the check source.

Since our target was bright (tens of mJy), we split the calibrated visibilities into smaller, roughly equal channel/frequency segments to better characterize the spectral shape. Specifically, the 2.1\,GHz and 5.5\,GHz bands were each divided into four segments, while the 9.0\,GHz band was divided into two segments. We performed imaging using the \texttt{Miriad} tasks \texttt{invert}, \texttt{restor}, and \texttt{mfclean}. We adopted a robust weighting of 0.5 as a compromise between sensitivity and sidelobe levels. We also performed a few rounds of phase self-calibration to correct for remaining phase errors as much as possible. The final Stokes I integrated flux densities were measured using the \texttt{imfit} task and are given in Table~\ref{tab:radioobs}. Note that the uncertainties were computed by adding in quadrature the fit error, the rms noise, and $5\%$ of the total flux density, the latter accounting for the uncertainty in the flux density scale of the primary calibrator. 

We applied the same imaging and self-calibration procedure on the check source NVSS\,J182359-235423 to assess the accuracy of our flux density measurements and any potential phase decorrelation that remains uncorrected. At lower frequencies ($\lesssim$\,3\,GHz), the measured ATCA flux densities of the check source were generally consistent with archival values from other radio surveys, including NVSS, VCSS, and VLASS. However, the check source lacks archival measurements at higher frequencies, so we were unable to assess the accuracy of our measurements in the 5.5\,GHz and 9.0\,GHz bands. Considering the agreement at lower frequencies and similar processing across all bands, we assume that phase decorrelation has not significantly impacted our flux density measurements.  

\begin{figure}
    \centering
    \includegraphics[width=0.45\textwidth]{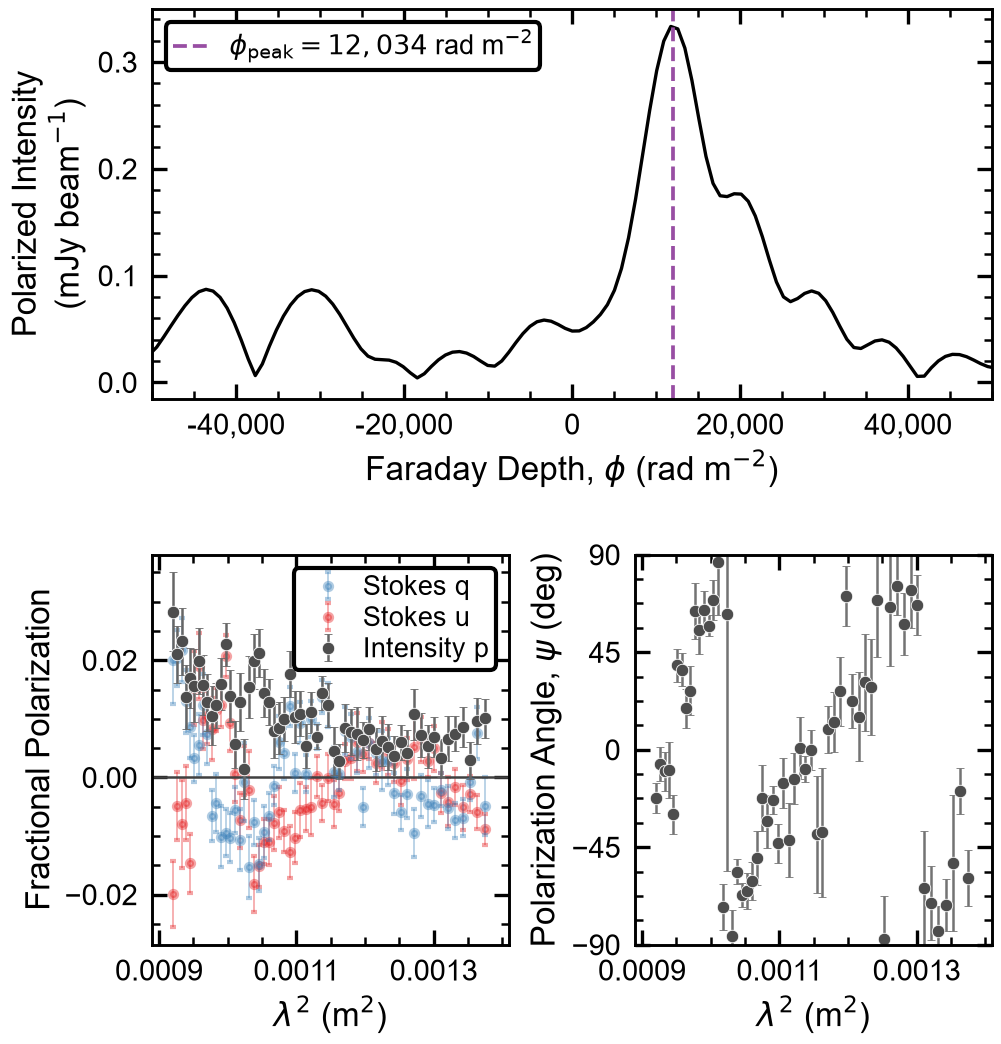}
    \caption{Resulting plots from RM synthesis for the ATCA 9.0\,GHz band. The top panel shows the Faraday dispersion function with a clear detection at $\phi = 12,034\,\mathrm{rad}\,\mathrm{m}^{-2}$. The bottom left and right panels show the fractional polarization and the polarization angle, respectively, both over $\lambda^2$.}
    \label{fig:RMsynth}
\end{figure}

In addition, we examined the Stokes Q, U, and V data to check for polarized emission. Starting with the 9.0\,GHz band, we generated Stokes I, Q, and U images averaged over every 32 channels (32\,MHz) and obtained spectra by measuring the peak flux density at the pixel location of VT\,J1823-2345 and the image rms at an off-source region. With the spectra, we performed Faraday rotation measure (RM) synthesis \citep[][]{BrentjensdeBruyn2005} using the \texttt{RMsynth1D} and \texttt{RMclean1D} algorithms in \texttt{RM-Tools} package \citep[V1.4.11;][]{VanEck2026}. Default parameters were used for \texttt{RMsynth1D}, but for \texttt{RMclean1D}, we used $10\sigma$ as the \textsc{clean} cutoff and $5\sigma$ for windowed \textsc{clean}. Some of the resulting plots are shown in Figure~\ref{fig:RMsynth}.

Based on RM synthesis, we found a significant linearly polarized component at an extreme Faraday depth of $\phi = (+12,034\pm357)\,\mathrm{rad}\,\mathrm{m}^{-2}$. Assuming a single Faraday-thin component, this implies $\mathrm{RM_{obs}}\sim+12,000\,\mathrm{rad}\,\mathrm{m}^{-2}$ in the observer frame. Strong Faraday rotation is clearly seen in the change in polarization angle. At the reference frequency of 8.8\,GHz, the linearly polarized intensity is about $0.33\,\mathrm{mJy}\,\mathrm{beam}^{-1}$, detected at a signal-to-noise ratio of 14.0. Note that this is the peak intensity from the RM synthesis that corresponds to the most prominent component of a potentially complex source. This yields a linear polarization fraction of $p_\mathrm{L} \simeq 0.73\%$.

For the 9.0\,GHz band, we also extracted the band-averaged Stokes V map and the Stokes V spectrum and found no detection of circularly polarized emission in both the image and the spectrum. Based on the image rms, the $5\sigma$ upper limit on the circular polarization fraction is $p_{\mathrm{C}}<0.2\%$ at 9.0\,GHz.

We also performed RM synthesis on the 5.5\,GHz band using images averaged over 16 channels (16\,MHz). We found the most significant detection to be near zero Faraday depth with a signal-to-noise ratio of 7.9 and $p_{\mathrm{L}}\simeq 0.34\%$ at a reference frequency of 5.5\,GHz. There is also a slightly weaker detection at $\phi \sim 11,194\,\mathrm{rad}\,\mathrm{m}^{-2}$ with a signal-to-noise ratio of 5.8 and $p_{\mathrm{L}}\simeq 0.25\%$. We suspect that the detection near zero Faraday depth is not a real signal and is caused by polarization leakage. We also checked the band-averaged Stokes V map at 5.5\,GHz and found a faint negative source with $p_{\mathrm{C}}\simeq0.34\%$, which likely indicates leakage since such circularly polarized emission was not detected at 9.0\,GHz.

On the other hand, the 5.5\,GHz detection at $\phi \sim 11,194\,\mathrm{rad}\,\mathrm{m}^{-2}$ is likely associated with the detection in the 9.0\,GHz band. However, since leakage dominates the signal and the signal-to-noise of this detection is relatively low, we treat it as an upper limit with $p_{\mathrm{L}}< 0.25\%$ at 5.5\,GHz. Nonetheless, compared to $p_\mathrm{L} \simeq 0.73\%$ at 8.8\,GHz, the much lower linear polarization fraction at 5.5\,GHz is an indication of depolarization. We note that the fractional depolarization caused by intra-channel Faraday rotation is only a few percentage over the 5.5\,GHz and 9.0\,GHz bands (considering our averaged channels). Therefore, the observed depolarization from 9.0\,GHz to 5.5\,GHz is likely intrinsic to VT\,J1823-2345.

Finally, we performed RM synthesis on the 2.1\,GHz band using images generated per channel (1\,MHz). We similarly recovered a detection near zero Faraday depth, which again is likely not a real signal. We found no detection near $\phi \sim 12,000\,\mathrm{rad}\,\mathrm{m}^{-2}$, which is unsurprising considering that significant depolarization is already seen at 5.5\,GHz.

\subsubsection{VLA and VLITE}

VT\,J1823-2345 was observed with the VLA on 2024 May 04 in the B configuration under project 24A-318 (PI: Y. Chen) to obtain another epoch of continuum measurements. The observations totaled 1.25\,hr over a single scheduling block that included L (1--2\,GHz), S (2--4\,GHz), C (4--8\,GHz), X (8-12\,GHz), and Ku-band (12--18\,GHz) observations. All observations were conducted in full polarization mode and the correlator was set up with 16 (L, S), 32 (C, X), or 48 (Ku) spectral windows and 64 channels per spectral window, where 1-MHz channels were used for L-band and 2-MHz channels were used for S, C, X, and Ku-band. For all bands, we observed 3C286 for flux density scale and bandpass calibration and the nearby calibrator 1817-254 for complex gain calibration. Pointing scans on the calibrators were also taken prior to the Ku-band and X-band observations.

We processed the VLA data using the Common Astronomy Software Applications \citep[CASA; version 6.6.1;][]{CASA} and used the VLA pipeline script\footnote{\url{https://science.nrao.edu/facilities/vla/data-processing/pipeline}} for calibration. Prior to processing, since our target was bright (tens of mJy), we split each band into individual segments containing either 8 (L, S) or 16 (C, X, Ku) spectral windows. For each segment, we first performed one calibration pipeline run and applied additional flagging on the product of the first run. On the flagged data, we then performed a second pipeline run but without deterministic flagging or Hanning smoothing. We used the calibrated data from the second run for imaging.

We performed imaging using the CASA task \texttt{tclean}, adopting the multi-term multi-frequency synthesis algorithm for deconvolution. A Briggs weighting \citep[][]{Briggs1995} of 0.5 was used to balance the sensitivity and sidelobe levels. For the lower-frequency L and S-bands, we excluded the shorter baselines $<$\,7\,k$\lambda$ during imaging to mitigate the impact of diffuse emission from the SNR. Lastly, we performed one to three rounds of phase self-calibration to further improve the final image quality. The final VLA measurements were obtained using the CASA \texttt{imfit} task, which are given in Table~\ref{tab:radioobs}. Note that the uncertainties also include 5\% of the flux density added in quadrature to the fit error to account for the uncertainty in the flux density scale calibration.

During the VLA observations at each of the five GHz bands, the commensal instrument VLITE also recorded data simultaneously at 340\,MHz, adding a low-frequency measurement. To process the VLITE data, each dataset was calibrated separately by the standard VLITE calibration and imaging pipeline \citep{Polisensky2016}, which uses tasks from the Obit \citep{Cotton2008} and the Astronomical Image Processing System (AIPS) software packages. The $(u,v)$-data were combined and self-calibrated in phase. The final Stokes I image was made using the Obit task \textsc{MFImage} with a robust weighting factor of zero. The data were restricted to baselines longer than $2$\,k$\lambda$ to minimize contributions from the diffuse structure of the SNR. The VLITE measurement of VT\,J1823-2345 at 340\,MHz is given in Table~\ref{tab:radioobs}, with an uncertainty that includes the fit error, the image rms, and a 10\% flux scaling uncertainty.

\begin{deluxetable}{cccc}
\tablecolumns{5}
\tablecaption{Radio Multi-band Measurements}
\tablehead{
Telescope &
Date       &
$\nu$  & 
$S_\nu$ \\
 & 
(UTC)   & 
(GHz) & 
(mJy)
}
\startdata
ATCA & 2023 Jan 21 & 1.63 & $36.0\pm 1.9$ \\
&    & 2.01 & $42.9\pm 2.2$ \\
&    & 2.43 & $50.8\pm 2.6$ \\
&    & 2.85 & $56.9\pm 2.9$ \\
&    & 4.78 & $70.3\pm 3.6$ \\
&    & 5.22 & $72.0\pm 3.6$ \\
&    & 5.72 & $72.4\pm 3.6$ \\
&    & 6.19 & $71.9\pm 3.6$ \\
&    & 8.52 & $64.0\pm 3.2$ \\
&    & 9.40 & $57.9\pm 2.9$ \\\hline
VLA/VLITE & 2024 May 04 & 0.34 & $22.4\pm6.1$ \\\hline
VLA & 2024 May 04 & 1.33 & $33.5\pm 1.7$ \\
&     & 1.81  & $41.0\pm 2.1$ \\
&     & 2.65  & $49.1\pm 2.5$ \\
&     & 3.49  & $56.7\pm 2.8$ \\
&     & 4.94 & $60.8\pm 3.0$ \\
&     & 7.07 & $55.2\pm 2.8$ \\
&     & 9.04 & $47.6\pm 2.4$ \\
&     & 10.93 & $42.6\pm 2.1$ \\
&     & 13.38 & $38.5\pm 1.9$ \\
&     & 14.85 & $36.9\pm 1.8$ \\
&     & 16.69 & $34.4\pm 1.7$ \\  
\enddata
\tablecomments{$1\sigma$ error is given after the $\pm$ sign.}
\label{tab:radioobs}
\end{deluxetable}

\subsubsection{VLBA}

VT\,J1823-2345 was observed by the Very Long Baseline Array (VLBA) on 2024 June 12 under project BC297 (PI Y. Chen) with a total observing time of 7.25\,hr. The observing session was conducted in the 2\,cm (or U) band with a total bandwidth of 512\,MHz in dual polarization mode. All stations participated in the observations except the Hancock station which was offline for generator installation. The data were correlated at the Socorro-DiFX correlator with a 1-second integration time.

The main objective of the VLBA observations was to measure an accurate radio position for comparison with the optical position of the coincident star. For this purpose, we employed both geodetic blocks and nodding style phase referencing using three nearby phase calibrators: J1822-2430, J1823-2240, and J1828-2417. Additionally, a fourth calibrator, J1820-2528, was observed as a phase check source to evaluate our final astrometric accuracy. The fringe finder J1924-2914 was observed for instrumental delay and bandpass calibration. 

We processed and analyzed the data using the AIPS 31DEC25 release. We followed the standard recipe for corrections and calibration. With the geodetic blocks, we also derived corrections for atmospheric delays. During fringe fitting, we found that two of the three phase calibrators were unfortunately too faint, which resulted in a large number of failed solutions. Therefore, we only performed fringe fitting on the brightest phase calibrator, J1822-2430, and applied the resulting solutions to our target VT\,J1823-2345 and the check source. 

With the calibrated data, we used the AIPS task \texttt{IMAGR} to perform imaging on the phase calibrator and our target. A robust weighting of 0 was adopted as a compromise between sensitivity and sidelobe levels. We excluded the North Liberty antenna during imaging after noticing phase instability, which resulted in improved beam size and peak intensity and a decrease in the image rms noise. 

\begin{figure}
    \epsscale{1.0}
    \centering
    \plotone{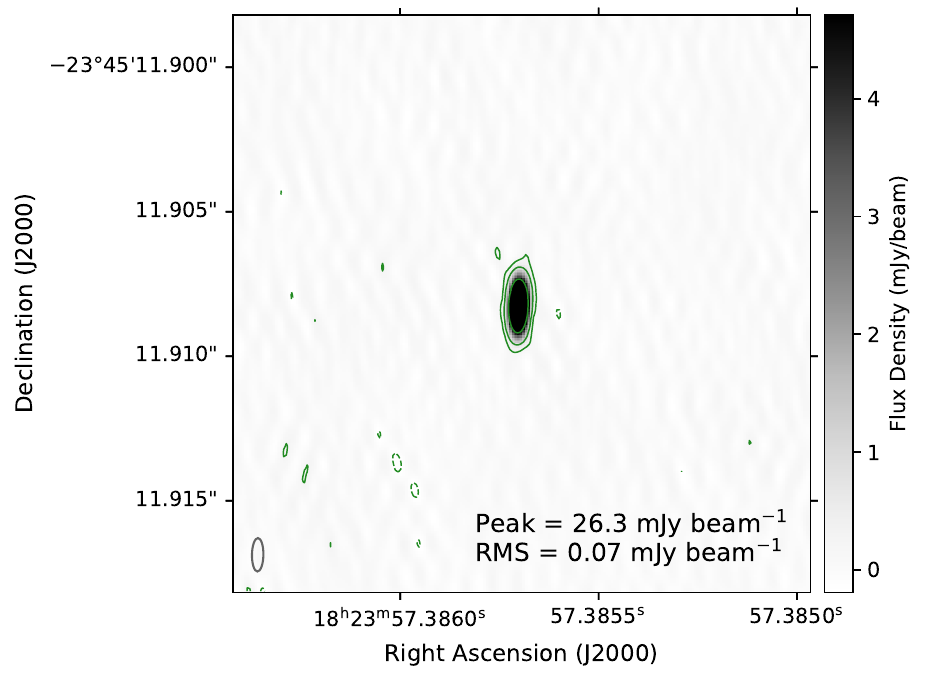}
    \caption{VLBA U-band (15\,GHz) image of VT\,J1823-2345. The restoring beam is shown at the bottom left with a size of $1.157\times0.396$\,mas and a position angle of $0.85^{\circ}$. The peak flux density and image rms are listed at the bottom right. The contours are $-$3, 3, 12, and 72 times the image rms, with negative contours shown with dashed lines.}
    \label{fig:VLBAJ1823img}
\end{figure}

We found that the phase calibrator J1822-2430 was resolved at 15\,GHz in a consistent manner with previous 2.3 and 8.4\,GHz observations \citep{Petrov2025RFC}. Therefore, to properly account for its structure, we performed non-cumulative rounds of self-calibration on the phase calibrator, first in phase only, and then in both amplitude and phase. These self-calibration solutions were then transferred to the target source. Imaging and self-calibration were then performed on the target source to produce the final VLBA image of VT\,J1823-2345, shown in Figure~\ref{fig:VLBAJ1823img}.

Using the AIPS task \texttt{JMFIT}, we fit an elliptical Gaussian model to the final image of our target VT\,J1823-2345, which yielded a peak intensity of $26.3\,\mathrm{mJy}\,\mathrm{beam^{-1}}$ and an integrated flux density of $S_\nu = 30.2\,\mathrm{mJy}$ at $\nu=15.17\,\mathrm{GHz}$. The uncertainty in the measured flux density is likely dominated by the uncertainty of the flux density scale that is typically $\gtrsim$\,10\%. Therefore, our VLBA measurement is broadly consistent with the VLA measurement in the Ku-band (Table~\ref{tab:radioobs}). 

The fit is also resolved, with a nominal deconvolved size of $\theta_{\mathrm{maj}}=0.40^{+0.01}_{-0.01}\,\mathrm{mas}$, $\theta_{\mathrm{min}}=0.08^{+0.02}_{-0.04}\,\mathrm{mas}$ in the major and minor axis, and a position angle of $(156\pm2)^{\circ}$. Note that the uncertainties are derived from the minimum and maximum values in the output of \texttt{JMFIT} (with default \textsc{efactor}\,$=$\,1.3). We also tried using different weighting (a robust weighting of -4 and 4) and recovered similar sizes. This finding appears consistent with the visibilities, which show amplitude structure on short baselines $\lesssim$\,50\,M$\lambda$ (see Appendix~\ref{apdx:VLBA} and Figure~\ref{fig:VLBAvis}). To double check, we also used the AIPS task \texttt{UVFIT} to fit an elliptical Gaussian model directly to the visibilities. \texttt{UVFIT} reported consistent results, with $S_\nu = 30.1\,\mathrm{mJy}$, $\theta_{\mathrm{maj}}=0.37\,\mathrm{mas}$, $\theta_{\mathrm{min}}=0.07\,\mathrm{mas}$, and a position angle of  $154^{\circ}$. To ensure that this fit is accurate, we generated the residual image from the model-subtracted visibilities after directly subtracting the \texttt{UVFIT} model from the calibrated visibilities. This residual image does not show significant emission or (error) structures.

To further test the modeling, we also tried fitting one or two circular Gaussian components to the visibilities using \texttt{UVFIT} and found that the circular model preferred a deconvolved FWHM of about 0.2\,mas in both cases. We also generated model-subtracted residual images and found significant emission and error structures (symmetric positive and negative components), indicating that the observed emission is not well-modeled by a single or a pair of circular Gaussian components. Therefore, our findings suggest that the elliptical Gaussian fit is likely reasonable, which potentially suggest an elongated structure along the fit position angle. For this study, we adopt the deconvolved sizes reported by \texttt{JMFIT} as the projected size of VT\,J1823-2345, though we note the caveat that a Gaussian model may not be fully representative of any complex physical structures.

Finally, we measured the VLBA position of VT\,J1823-2345 using an image constructed without self-calibration or any transfer of self-calibration solutions. We found the position to be at R.A.(J2000)\,=\,18h23m57.3857s and Decl.(J2000)\,=\,-23d45m11.9082s, which is 0.914\,arcsec away from the Gaia DR3 position of the star at the time of the VLBA observations after accounting for proper motion. We note that the fitting errors in the position is extremely small (on the order of $\mu$as) and does not reflect the actual uncertainty in the phase referencing. Therefore, to assess the accuracy of the astrometry, we similarly measured the position of our check source J1820-2528 using an image constructed without any self-calibration. We found our check source in the U-band to be offset from its ICRF3-SX \citep[][]{Charlot2020ICRF3SX} position by 0.28\,mas in R.A. and 0.30\,mas in Decl. These offsets are $\sim$\,$2\sigma-4\sigma$ compared to the formal ICRF3-SX uncertainties but are consistent with the excess astrometric variability of this check source \citep[$\sigma_{\mathrm{R.A.}}\approx0.3\,\mathrm{mas}$ and $\sigma_{\mathrm{Decl.}}\approx0.6\,\mathrm{mas}$;][]{Cigan2024}.
Therefore, our measured position of VT\,J1823-2345 should be accurate to within $\lesssim$\,mas precision, which is more than sufficient for confirming a separation of 0.914\,arcsec. The radio sizes and position from our VLBA observations are listed in Table~\ref{tab:radioproperties}. 

During the writing of this manuscript, we also found additional VLBA data for VT\,J1823-2345 posted on the Astrogeo VLBI FITS image database\footnote{\url{https://astrogeo.org/vlbi_images/}}. This data was taken by the VLBA on 2026 August 23 under project BP280 (PI L. Petrov) in the C and X band with a scan time of 54 seconds. The database now lists this source (J1823-2345) as a VLBI calibrator \citep[][]{Petrov2025RFC} and hosts the images and visibilities from the 2026 snapshot. Here, we briefly describe the data products downloaded directly from the database and results from some additional analyses.

In 2026, VT\,J1823-2345 still appears as a single component in the VLBA images. The 2026 VLBA position of VT\,J1823-2345 listed on the Astrogeo database is consistent with our 2024 VLBA position to within uncertainty. The visibilities show evidence of declining amplitude over $(u,v)$ distance for C band, but this decline appears less significant in X band. Based on the flux densities reported on the Astrogeo database, the unresolved flux density accounts for about 78\% of the total in C band and about 90\% of the total in X band. With evidence of resolved structure, we performed additional analyses on the images and visibilities to characterize the emission.

Using \texttt{JMFIT} to fit a single elliptical Gaussian component to the 2026 images, we recovered an integrated flux density of $S_\nu = 38.0\,\mathrm{mJy}$ at $\nu=4.30\,\mathrm{GHz}$ in C band and $S_\nu = 28.4\,\mathrm{mJy}$ at $\nu=7.57\,\mathrm{GHz}$ in X band, which demonstrate continuous fading at these frequencies compared to our ATCA and VLA measurements taken a few years prior (Table~\ref{tab:radioobs}). The spectral index between these two frequencies appears to be steepening (potentially implying a shift in the peak), but this finding is not constrained considering the large uncertainty in the VLBA flux density scale.

To constrain the emission size, we examined the elliptical Gaussian fit from both \texttt{JMFIT} and \texttt{UVFIT} and found a consistent deconvolved major axis size of $\theta_{\mathrm{maj}}\sim0.6-0.7\,\mathrm{mas}$ in both C and X band, but the minor axis size and position angle were generally inconsistent between different fits and different bands. This is most likely the result of the short scan time and limited $(u,v)$-coverage, so we also tried limiting the fit by using a circular Gaussian and only fitting the visibilities. With a circular Gaussian, we found a FWHM of $0.60\pm0.04\,\mathrm{mas}$ for C band and $0.28\pm0.05\,\mathrm{mas}$ for X band from \texttt{UVFIT}. 
Overall, our findings suggest that VT\,J1823-2345 is also resolved in the 2026 VLBA data at lower frequencies with sizes that are comparable to (perhaps slightly larger than) the 2024 sizes, but the limited $(u,v)$-coverage prevents us from making any detailed interpretations of the source structure. Therefore, in this study, we do not attempt to examine any potential source evolution or optical depth effects with the 2026 data.

\subsubsection{Archival and Survey Radio Measurements}

In addition to our multi-band follow-up observations, we also collected a number of archival and survey measurements in the radio to broadly examine the temporal evolution of VT\,J1823-2345. At 3.0\,GHz, we obtained VLASS epoch 1 and 2 flux densities from the quick look (QL) component catalogs produced by the Canadian Initiative for Radio Astronomy Data Analysis\footnote{See \url{https://cirada.ca/vlasscatalogueql0}.} \citep[CIRADA;][]{Gordon2021VLASS}{}{}. We also measured VLASS epoch 3 flux density from the released QL image using CASA task \texttt{imfit}, which yielded $S_\nu = (52.5\pm2.7)\,\mathrm{mJy}$. At 1.4\,GHz, we collected measurements from NVSS and RACS-mid1. At 887.5\,MHz, we obtained a set of measurements from the first two epochs of the RACS low-band survey \citep[][]{McConnell2020RACSI,Hale2021RACSIIcatalog} and VAST \citep[][]{Murphy2021VAST,deRuiter2026VASTextDR1}. With the exception of NVSS, these radio survey measurements span a time range of $\sim$\,2019--2024. Note that systematic uncertainties were also added in quadrature for these measurements to account for the flux scaling uncertainties.

\begin{figure*}
    \epsscale{1.0}
    \centering
    \plotone{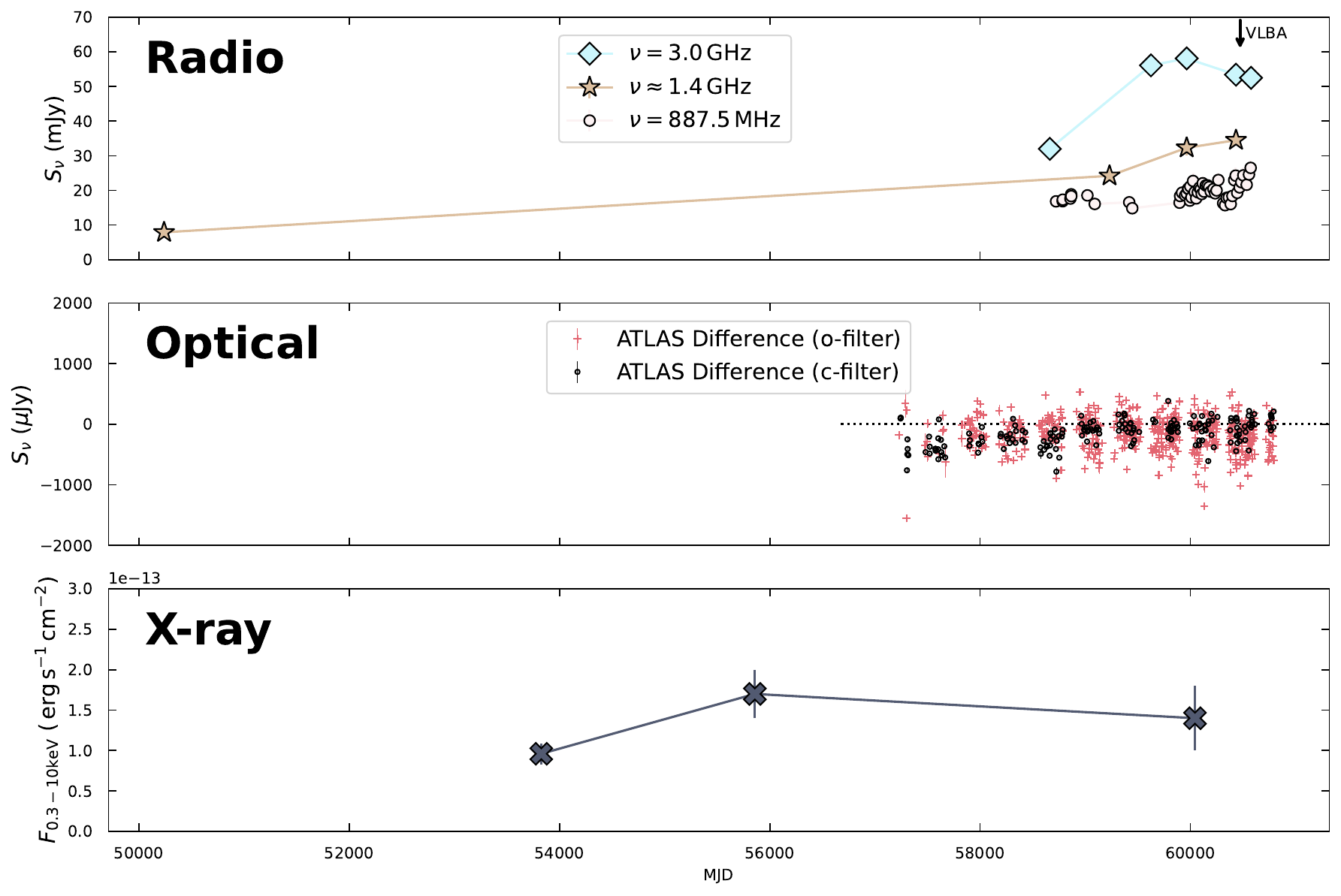}
    \caption{Light curves of VT\,J1823-2345 in the radio (top panel), optical (middle panel), and X-ray (bottom panel). Frequencies are labeled for the radio light curves, and an arrow shows the time of the VLBA observations. The optical light curves contain forced photometry from ATLAS difference images, and a dashed line is shown at zero flux density. The X-ray (unabsorbed) fluxes are all measured over 0.3--10\,keV assuming a power law with $\Gamma_\mathrm{X} = 1.7$ and a Galactic absorption of $N_{\mathrm{H}}=2.07\times10^{21}\,\mathrm{cm}^{-2}$.}
    \label{fig:LCs}
\end{figure*}

The collected radio light curves of VT\,J1823-2345 are shown at the top panel of Figure~\ref{fig:LCs}. Note that the light curves include points interpolated/extrapolated from the best-fit radio spectra for the multi-band ATCA and VLA observations (see Section~\ref{subsubsec:radSED}). For example, the 3.0\,GHz light curve contains three points from VLASS and two points from ATCA and VLA. VT\,J1823-2345 appears to be evolving smoothly over years, with the 3.0\,GHz light curve peaking at around 2023 while the lower-frequency emission continues to brighten (also see Figure~\ref{fig:RadioLCs}).

\subsection{X-ray Observations}

\subsubsection{Swift}

The location of VT\,J1823-2345 was observed serendipitously by the Neil Gehrels Swift Observatory (Swift) in 2011 in the field of GRB\,111022A (target id 506070). There were four observations in 2011 from October 22 to October 25 with a total exposure time of $\sim$\,18\,ks in the X-Ray Telescope (XRT) in photon counting (PC) mode. We also obtained two follow-up Swift observations of VT\,J1823-2345 (target id 15947) in 2023 from April 8 to April 13 with a total exposure time of $\sim$\,9\,ks in the XRT in PC mode. 

To analyze the Swift XRT observations, we downloaded the cleaned event files and the vignetting-corrected exposure maps from the Swift data archive. We used the HEASoft \citep[version 6.35.1;][]{HEAsoft} tool \textsc{xselect} to extract images from the event files and \textsc{ximage} to combine and analyze the images. Specifically, we combined individual images and exposure maps to produce a single image and exposure map for both 2011 and 2023. 

We were able to extract a faint X-ray source at the location of VT\,J1823-2345 in both the 2011 and 2023 images, and we used the \texttt{sosta} command on the exposure-corrected image to derive the source count rate over 0.3--10\,keV. We found the source count rate to be $(3.00 \pm 0.54)\times10^{-3}\,\mathrm{cts}\,\mathrm{s}^{-1}$ in 2011 and $(2.42 \pm 0.68)\times10^{-3}\,\mathrm{cts}\,\mathrm{s}^{-1}$ in 2023. Both detections are statistically significant, with the probability of the observed counts arising from background fluctuation being $\sim$\,$6\times10^{-16}$ in 2011 and $\sim$\,$10^{-8}$ in 2023, assuming Poisson statistics. 

Our results suggest that VT\,J1823-2345 had approximately the same X-ray brightness in 2011 and 2023. With the count rates, we used PIMMS\footnote{\url{https://cxc.harvard.edu/toolkit/pimms.jsp}} (version 4.14) to estimate the corresponding X-ray flux. Assuming a power index of $\Gamma_{\mathrm{X}} = 1.7$ (standard non-thermal spectrum), a Galactic neutral hydrogen column density of $N_{\mathrm{H}}=2.07\times10^{21}\,\mathrm{cm}^{-2}$ \citep{HI4PI2016}, no redshift, and no intrinsic absorption, we obtained an unabsorbed flux estimate of $F_{0.3-10\mathrm{keV}}\sim(1.7\pm0.3)\times10^{-13}\,\mathrm{erg}\,\mathrm{s}^{-1}\,\mathrm{cm}^{-2}$ in 2011 and $F_{0.3-10\mathrm{keV}}\sim(1.4\pm0.4)\times10^{-13}\,\mathrm{erg}\,\mathrm{s}^{-1}\,\mathrm{cm}^{-2}$ in 2023. The absorbed fluxes are  $F_{0.3-10\mathrm{keV}}\sim(1.4\pm0.3)\times10^{-13}\,\mathrm{erg}\,\mathrm{s}^{-1}\,\mathrm{cm}^{-2}$ in 2011 and $F_{0.3-10\mathrm{keV}}\sim(1.1\pm0.3)\times10^{-13}\,\mathrm{erg}\,\mathrm{s}^{-1}\,\mathrm{cm}^{-2}$ in 2023.

\subsubsection{XMM-Newton}

The location of VT\,J1823-2345 was also observed by the XMM-Newton European Photon Imaging Camera (EPIC) MOS and pn in 2006 on April 1 during a $\sim$\,12\,ks observation targeting the SNR G8.7-5.0 (Obs. ID 0306700201; PI R. Bandiera). We downloaded the data from the XMM-Newton Science Archive and performed data reduction using the XMM-Newton Science Analysis System (SAS version 22.1). We used the tasks \texttt{epchain} and \texttt{emchain} to generate MOS1, MOS2, and pn event files. We then used the \texttt{espfilt} task to filter time intervals impacted by high background flares. 

When generating images, we selected single, double, triple, and quadruple events for MOS1 and MOS2 (PATTERN\,$\leq$\,12) and only single and double events for pn (PATTERN\,$\leq$\,4). For source detection, we ran the \texttt{edetect\_chain} task on the MOS1, MOS2, and pn images simultaneously. We identified a coincident X-ray source located 0.42\,arcsec from the VLBA radio position of VT\,J1823-2345 that has a positional uncertainty of 0.63\,arcsec. This X-ray source has a net count of $83\pm11$ in pn, $45\pm8$ in MOS1, and $52\pm8$ in MOS2 over the 0.3--10\,keV range with a likelihood of detection (DET\_ML) of 229.8 (corresponding to a probability of $10^{-99.8}$ that the detection is caused by random Poisson fluctuation). The corresponding count rate is $(2.2\pm0.3)\times10^{-2}\,\mathrm{cts}\,\mathrm{s}^{-1}$ in pn, $(0.8\pm0.1)\times10^{-2}\,\mathrm{cts}\,\mathrm{s}^{-1}$ in MOS1, and $(0.9\pm0.1)\times10^{-2}\,\mathrm{cts}\,\mathrm{s}^{-1}$ in MOS2.

With the pn count rate and the same assumptions of an absorbed power law with $\Gamma_{\mathrm{X}} = 1.7$ and $N_{\mathrm{H}}=2.07\times10^{21}\,\mathrm{cm}^{-2}$, we derived using PIMMS an unabsorbed X-ray flux estimate of $F_{0.3-10\mathrm{keV}}\sim(0.96\pm0.13)\times10^{-13}\,\mathrm{erg}\,\mathrm{s}^{-1}\,\mathrm{cm}^{-2}$ in 2006, or an absorbed flux of $F_{0.3-10\mathrm{keV}}\sim(0.79\pm0.11)\times10^{-13}\,\mathrm{erg}\,\mathrm{s}^{-1}\,\mathrm{cm}^{-2}$. This flux is broadly consistent with the Swift X-ray flux measured in 2011 and 2023 at the location of VT\,J1823-2345. The three X-ray measurements are shown in Figure~\ref{fig:LCs}.

We also extracted spectra from the filtered MOS1, MOS2, and pn event files to check the rough X-ray spectral shape. We used a 20\,arcsec circle centered on the X-ray source to extract the source spectrum and a nearby 100\,arcsec circle on the same CCD to extract the background spectrum. After generating the response and effective area (rmf and arf) files, we binned the spectra using the \texttt{specgroup} task to achieve a minimum signal-to-noise of 1.0 per bin. We fit the three spectra simultaneously over 0.3--10\,keV using \textsc{xspec} and employed W statistic due to the low count numbers. We modeled the spectra using an absorbed power law (\texttt{tbabs*powerlaw}) with the absorber fixed to the Galactic $N_{\mathrm{H}}=2.07\times10^{21}\,\mathrm{cm}^{-2}$ and found a best-fit photon index of $\Gamma_{\mathrm{X}} = 1.5^{+0.3}_{-0.3}$ (range corresponding to the 90\% confidence interval) with \texttt{goodness} returning 0\%. Although the uncertainty is large, this result confirms that our previous assumption of a non-thermal power law spectrum is likely appropriate. 

\subsection{Optical and Infrared Observations}

\subsubsection{Optical Spectra of Coincident Star}

We obtained optical spectra of the coincident F-type star using the Wide-Field Spectrograph (WiFeS). WiFeS is an Integral Field Unit mounted on the Australian National University 2.3\,m Telescope \citep[][]{Dopita2007}, located at Siding Spring Observatory near Coonabarabran, Australia. A series of WiFeS spectra was obtained between April and June 2023 to search for variations in radial velocity (i.e., binary motion) and signatures of an accreting compact object (e.g., an X-ray binary). The WiFeS spectra were taken in ``Nod \& Shuffle" mode, which produced simultaneous science and sky spectra, and sky contribution is subtracted during reduction \citep[see Section 2.2 of][for more details]{Carr2024}. The observations were reduced using the default reduction pipeline pyWiFeS \citep{Childress2014} to produce calibrated 3D data cubes, from which 2D spectra were extracted. Note that the extraction size is similar to the seeing, averaging between about 2\,--\,3\,arcsec, so the spectra could cover potential optical emission from the radio position.

We also obtained a spectrum on 2024 June 8 using the Inamori Magellan Areal Camera and Spectrograph \citep[IMACS;][]{Dressler2011IMACS} on the Magellan Baade telescope at the Las Campanas Observatory. The spectrum was taken with the f/4 camera of IMACS using a multislit slit mask consisting of 1.0''$\times$6.0'' slits, one of which was centered on the F-type star. Data were taken at two grating tilts with the 600\,lines\,mm$^{-1}$ grating, along with contemporaneous dome arcs and flats. Data reduction was performed using the COSMOS-3 reduction pipeline \citep[][]{Oemler2017COSMOS}, which involved standard mask alignment and rectification, flat fielding, sky subtraction, and extraction. These spectra are discussed in Appendix~\ref{apdx:OptSpec} and are presented in Figure~\ref{fig:FstarSpecSED}. In summary, we find that the optical spectra are consistent with an F2V star with no measured changes in radial velocity and no signs of excess emission. Therefore, we have no evidence that the F-type star is in a binary with an active compact object.

\subsubsection{Archival Optical and Infrared Photometry}

We searched various data archives and found that the Panoramic Survey Telescope and Rapid Response System 1 \citep[Pan-STARRS1 or PS1;][]{Chambers2016PS1} took the highest angular resolution broadband optical images at the location of VT\,J1823-2345. As an example, in Figure~\ref{fig:radoptimg}, we show a three-color image made from the stacked ($g$, $r$, and $i$-band) PS1 images. Evidently, the image at the location of VT\,J1823-2345 is dominated by the nearby F-type star, which has an AB magnitude of $\sim$\,14--13 from $g$-band to $y$-band. In our later analysis (Section~\ref{subsec:optassoc}), we will use the PS1 warp image from an individual observation to examine constraints on possible optical sources that could be behind the F-type star at the location of VT\,J1823-2345 (Section~\ref{subsec:optassoc}). 

We also obtained forced photometry at the VLBA position of VT\,J1823-2345 from difference images of the Asteroid Terrestrial-impact Last Alert System \citep[ATLAS;][]{Tonry2018ATLAS,Smith2020ATLASTransient,Shingles2021ATLASWeb}. The resulting ATLAS light curves in the cyan and orange filters, spanning $\sim$\,2015--2025, are shown in Figure~\ref{fig:LCs} after removing and clipping bad points. The ATLAS light curves do not show any significant variability over this time range, and the forced photometry data scatter roughly around zero. We double checked a number of the difference images and found that the scatter and the slight overabundance of negative values are the result of imperfect image subtraction.

In addition, we collected multi-band optical and infrared (IR) photometry from PS1, the Two Micron All Sky Survey \citep[2MASS;][]{Skrutskie20062MASS}, and the Wide-field Infrared Survey Explorer \citep[WISE;][]{Wright2010WISE} to compare the observed photometry with the expected magnitudes of the F-type star. The details are described in Appendix~\ref{apdx:OptSpec}, but in summary, we find the observed photometry to be broadly consistent with the expected magnitudes of the F-type star. The archival optical and IR photometry do not show strong evidence of excess emission from a background object at the radio position. We also checked the PS1 and WISE light curves and found no significant fluctuations or flares above the F-type star. 

\section{Analysis and Results} \label{sec:analysis}

In this section, we analyze and characterize the observed emission features of VT\,J1823-2345, including the radio properties and the optical emission profile at the radio position.

\subsection{Radio Properties}

\subsubsection{Light Curve} \label{subsubsec:radLC}

\begin{figure}
    \epsscale{1.0}
    \centering
    \plotone{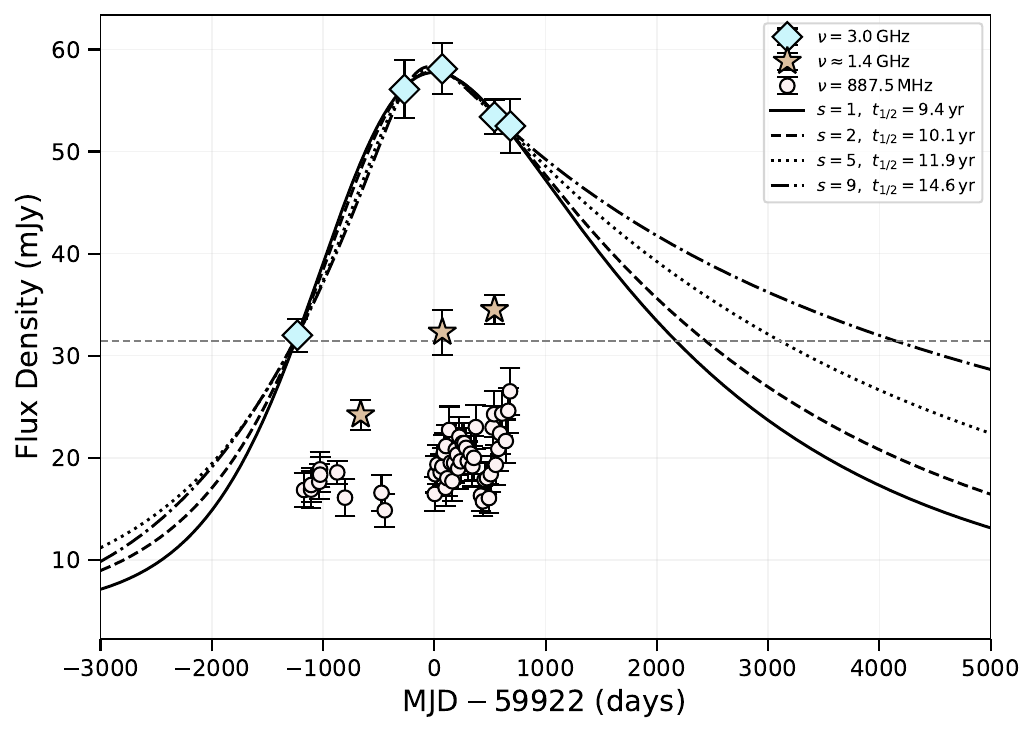}
    \caption{Radio light curves of VT\,J1823-2345 centered at MJD 59922 (the approximate peak of the 3.0\,GHz light curve). Best-fit smoothly broken power laws to the 3.0\,GHz light curve are also shown (solid and dashed curves), assuming different values of smoothness $s$. The time above half-maximum is also labeled with the best-fit curves, and the approximate half-maximum point is shown as a horizontal dashed line.}
    \label{fig:RadioLCs}
\end{figure}

In Figure~\ref{fig:RadioLCs}, we show the radio light curves around the peak time of the 3.0\,GHz light curve. To characterize the brightening and fading at 3.0\,GHz, we fit the light curve using a smoothly broken power law in the form:
\begin{equation}
    S_\nu(t)=S_{\nu,0} + A\left( y(t)^{-\beta_1 s} + y(t)^{\beta_2 s} \right)^{-1/s},
\end{equation}
where $S_\nu$ is the flux density, $A$ is the amplitude, $\beta_1$ and $\beta_2$ are the power law indices, $s$ is the smoothness parameter, and $y(t)=(t-t_0)/t_b$ with $t_0$ being the start time and $t_b$ being the break time. In this form, the peak or turnover time is: 
\begin{equation}
    t_{\mathrm{p}} = t_0+t_b\left( \frac{\beta_1}{\beta_2} \right)^{\frac{1}{s(\beta_1+\beta_2)}}.
\end{equation}
Since our light curve is sparse, we made a number of assumptions for the fit. We adopted $S_{\nu,0}=5\,\mathrm{mJy}$ as the stable host brightness, which is slightly lower than the 1.4\,GHz measurement from NVSS (assuming a standard negative spectral slope). We kept $t_0$ as a free parameter, but restricted the earliest start time to the NVSS observing time in 1996 (assuming the brightening began after the NVSS observations). We also fixed the smoothness parameter $s$ and examined the fit for various values of $s$. The \texttt{least\_squares} module of SciPy was used for this fit. We show the best-fit curves for four different smoothnesses in Figure~\ref{fig:RadioLCs}. 

The best-fit curves accurately describe the measurements and suggest that $t_p$ occurs at around MJD\,59900. Unfortunately, even with the above assumptions, we are unable to obtain reasonable values for $t_0$, $\beta_1$, or $\beta_2$ because our measurements provide very little constraints on the start time, and thus we do not report the best-fit parameters. Instead, we report the time above half-maximum $t_{1/2,\mathrm{obs}}$ from the best-fit curves because the observations (especially the rising portion) provide much stronger constraints on this parameter. The value of $t_{1/2,\mathrm{obs}}$ changes depending on the assumed smoothness (Figure~\ref{fig:RadioLCs}), but we find an order-of-magnitude estimate of $t_{1/2,\mathrm{obs}}\gtrsim10\,\mathrm{yr}$. This suggests that VT\,J1823-2345 is evolving over a characteristic timescale of about a decade, and the total duration could span many decades (although this timescale may be shorter if VT\,J1823-2345 is extragalactic and the observed timescale is dilated). 

Compared to the 3.0\,GHz light curve, the 1.4\,GHz and 887.5\,MHz light curves appear to still be brightening. The 887.5\,MHz light curve is much better sampled and shows potential evidence of fluctuation by a few tens of percent. For the 887.5\,MHz light curve, we find a $\chi^2$ value of 88.7 for a model of stable light curve \citep[following Equation 9 of][]{MurphyKaplan2026} with $N=52$ data points. Assuming a $\chi^2$ distribution with $N-1$ degrees of freedom, we find a probability of $0.084\%$ that $\chi^2$ exceeds the observed value due to random fluctuation, so variability in this light curve is statistically probable. However, we note that the level of fluctuation observed could be explained by interstellar scintillation, particularly at low frequencies \citep[][]{Walker1998}. Therefore, we do not attempt to associate this fluctuation with physical structures in the transient.

\subsubsection{Spectral Energy Distribution} \label{subsubsec:radSED}

\begin{figure}
    \epsscale{1.0}
    \centering
    \plotone{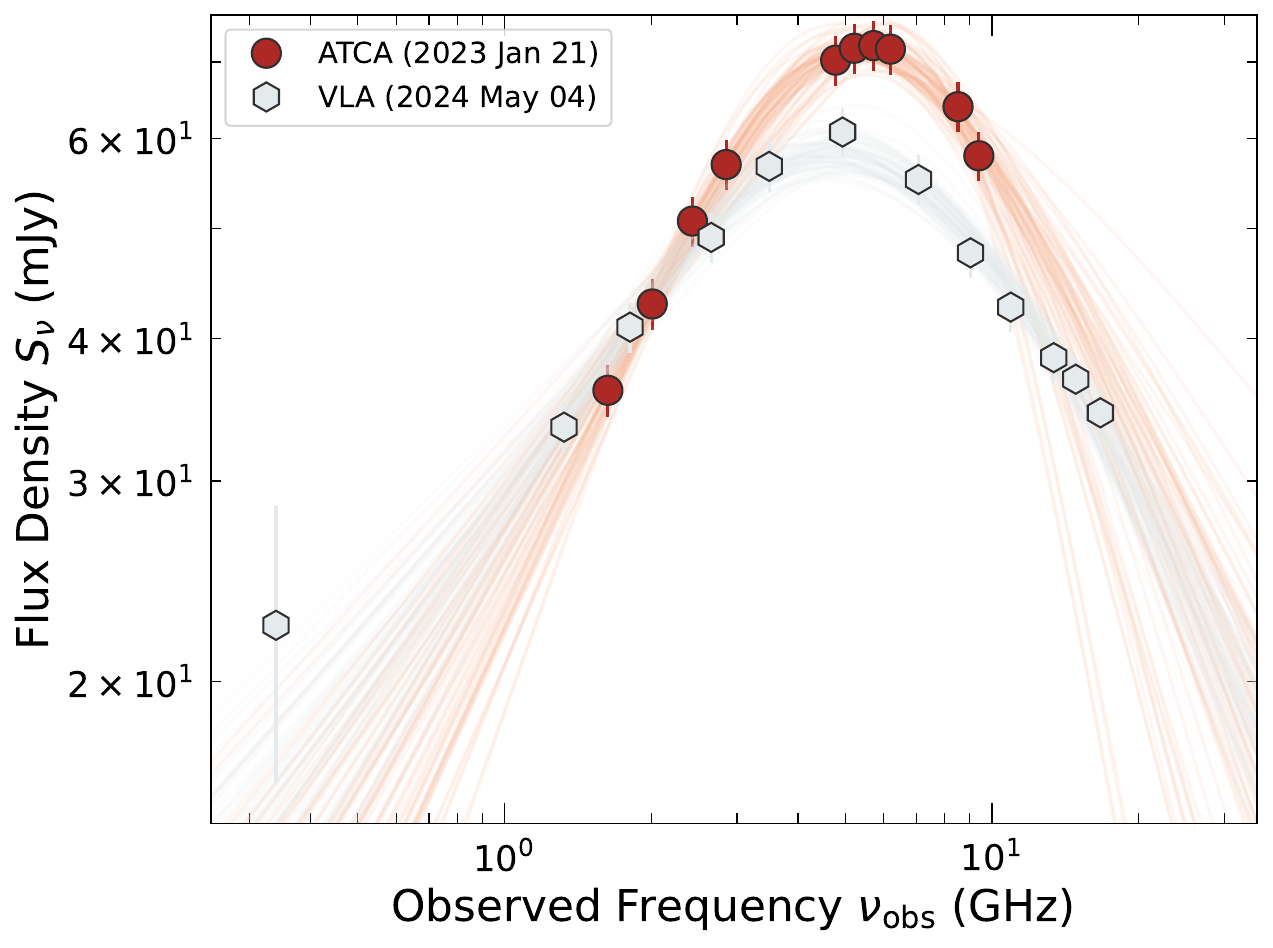}
    \caption{Radio SEDs measured with the ATCA and VLA (with VLITE) along with sample curve models from the MCMC fits. }
    \label{fig:RadioSEDs}
\end{figure}

Our multi-band radio measurements show a clear peaked spectrum over $\sim$\,1--10\,GHz (Figure~\ref{fig:RadioSEDs}). To characterize the spectrum, we used the Python \texttt{emcee} \citep{emcee2013} Markov Chain Monte Carlo (MCMC) sampler to fit the observed radio spectral energy distributions (SED). We used a Gaussian likelihood function and adopted a generic curved model of the form:
\begin{equation}
    S_{\mathrm{\nu}} = \frac{S_{\mathrm{\nu,p}}}{1-e^{-1}} \left( 1 - e^{-(\nu/\nu_{\mathrm{p}})^{\alpha_{\mathrm{thin}}-\alpha_{\mathrm{thick}}}} \right)\left( \frac{\nu}{\nu_{\mathrm{p}}} \right)^{\alpha_{\mathrm{thick}}},
\end{equation}
where $S_{\mathrm{\nu,p}}$ is the flux density at the peak frequency $\nu_{\mathrm{p}}$, and $\alpha_{\mathrm{thin}}$ and $\alpha_{\mathrm{thick}}$ are the spectral indices at the optically thin and optically thick regimes, respectively. The sampler was initialized with 128 walkers and run for an initial 1000 burn-in steps and a subsequent 2000 steps. The best-fit model parameters, $S_{\mathrm{\nu,p}}$, $\nu_{\mathrm{p,obs}}$, $\alpha_{\mathrm{thin}}$, and $\alpha_{\mathrm{thick}}$, were drawn from the 50th percentile of the posterior distributions, while the uncertainties were drawn from the 16th and 84th percentile. Note that $\nu_{\mathrm{p,obs}}$ denotes the peak frequency in the observer frame. These best-fit values are tabulated in Table~\ref{tab:radioproperties}. Some sample fits are shown in Figure~\ref{fig:RadioSEDs}.

Our fits show that between the ATCA and VLA observations (a span of 469 days in the observer frame), the spectrum of VT\,J1823-2345 is slowly evolving at $\sim$GHz frequencies with the peak brightness decreasing over time. The peak frequency may have decreased between the ATCA and VLA observations, but we do not measure any statistically significant changes in $\nu_\mathrm{p}$. It is interesting to note that the optically thick spectral index $\alpha_{\mathrm{thick}}\sim0.6$--$0.8$ is fairly shallow compared to the expected index of 2.0 or 2.5 for synchrotron self-absorption or free-free absorption in a homogeneous environment.

We also note that SNR\,G8.7-5.0 was recently observed by the MeerKAT telescope \citep[][]{JonasMeerKAT2016} in 2022 May 22 \citep[][]{Cotton2024} at $\nu\approx 1.3\,\mathrm{GHz}$, a few months before our ATCA observation. The published MeerKAT spectral index map shows $\alpha\sim+0.7$ at the location of VT\,J1823-2345, consistent with our ATCA and VLA observations. Therefore, the spectral shape of VT\,J1823-2345 does not appear to show major or sporadic changes over time, at least over the inverted (optically thick) portion.

\begin{deluxetable}{ccc}
\tablecolumns{3}
\tablecaption{Radio Properties of VT\,J1823-2345}
\tablehead{
Date       &
Property  & 
Value \\
(UTC)   & 
 &
}
\startdata
2023 Jan 21 & $\nu_{\mathrm{p,obs}}$ (GHz) & $6.65^{+1.23}_{-1.14}$ \\
& $S_{\nu,\mathrm{p}}$ (mJy) & $69.38^{+2.43}_{-3.52}$ \\
& $\alpha_{\mathrm{thick}}$ & $0.79^{+0.17}_{-0.14}$ \\
& $\alpha_{\mathrm{thin}}$ & $-1.08^{+0.25}_{-0.36}$ \\\hline
2024 May 04 & $\nu_{\mathrm{p,obs}}$ (GHz) & $6.64^{+1.03}_{-1.03}$ \\
& $S_{\nu,\mathrm{p}}$ (mJy) & $55.30^{+2.15}_{-2.39}$ \\
& $\alpha_{\mathrm{thick}}$ & $0.59^{+0.11}_{-0.09}$ \\
& $\alpha_{\mathrm{thin}}$ & $-0.91^{+0.11}_{-0.12}$ \\\hline
2024 Jun 12 & R.A. (J2000) & 18h23m57.3857s\\
& Decl. (J2000) & -23d45m11.9082s\\
& $\theta_{\mathrm{maj}}$ (mas) & $0.40^{+0.01}_{-0.01}$ \\
& $\theta_{\mathrm{min}}$ (mas) & $0.08^{+0.02}_{-0.04}$ \\
& Position Angle ($^{\circ}$) & $156\pm2$ \\
\enddata
\tablecomments{Errors are from the 16th and 84th percentiles. Radio position (J2000) is accurate to at least $<$\,1\,mas.}
\label{tab:radioproperties}
\end{deluxetable}

\subsubsection{Brightness Temperature}

Based on the the angular size obtained from the VLBA observations, we can estimate the brightness temperature $T_{\mathrm{B}}$ of VT\,J1823-2345. We calculate $T_{\mathrm{B}}$ using the formula \citep[][]{KunertBajraszewska2025}:
\begin{eqnarray}
    \frac{T_{\mathrm{B}}}{1+z} &=& 1.38\times10^{12}\,\mathrm{K}\left( \frac{S_{\nu}}{\mathrm{Jy}} \right)\left( \frac{\nu}{\mathrm{GHz}} \right)^{-2}\left(\frac{\theta_{\mathrm{}}}{\mathrm{mas}} \right)^{-2},
\end{eqnarray}
where $z$ is the redshift, $\theta$ is the angular size, and $S_\nu$ is the flux density measured at the observed frequency $\nu$. Taking $S_\nu = 30.2\,\mathrm{mJy}$, $\nu=15.17\,\mathrm{GHz}$, and $\theta^2 = \theta_{\mathrm{maj}}\times\theta_{\mathrm{min}}$, we find $T_{\mathrm{B}}/(1+z) = 5.7\times10^{9}\,\mathrm{K}$. Such brightness temperature indicates non-thermal emission and is consistent with brightness temperatures found in X-ray binaries, supernovae (SNe), and AGN in radio galaxies \citep[e.g.,][]{Pietka2015,Morabito2022}.

\subsection{Constraints on Optical Association} \label{subsec:optassoc}

\begin{figure*}
    \epsscale{1.0}
    \centering
    \plotone{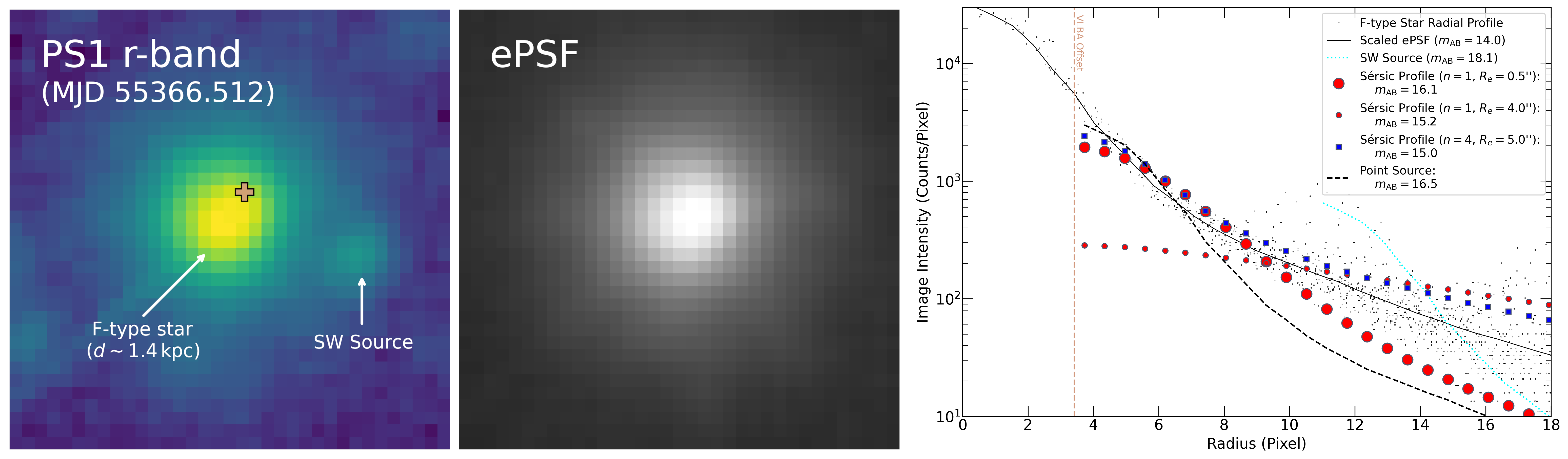}
    \caption{\textit{Left}: PS1 $r$-band warp image with a cross indicating the VLBA position of VT\,J1823-2345. \textit{Middle}: Image of the ePSF of the PS1 $r$-band warp image. \textit{Right}: Comparisons of radial profiles of the F-type star, scaled ePSFs, and hypothetical foreground or background objects at the VLBA position of VT\,J1823-2345.}
    \label{fig:PS1RadProf}
\end{figure*}

In the case that VT\,J1823-2345 originated from a background object or a foreground active star, we can still obtain some constraints on this object by examining the distribution of optical emission around the position of VT\,J1823-2345. To do this, we selected a PS1 $r$-band warp image from one individual observation \citep[][]{Chambers2016PS1} rather than using the stacked image that combines all observations. The reason for this choice is that the stacked image shows a broader spread in the emission of the F-type star due to the inclusion of images with poorer seeing.

For the selection, we retrieved warp images in which the location of VT\,J1823-2345 did not land on a gap and was not masked for any reason in the mask image. We then examined the image headers and selected one with a relatively small seeing FWHM and a reasonably large zero point. Ultimately, we selected an $r$-band image taken at MJD\,55366.512 (left of Figure~\ref{fig:PS1RadProf}) with a chip seeing FWHM of 3.89\,pix or 0.97\,arcsec (the scale being 0.25\,arcsec/pix). With this image and the mask image, we used the \texttt{Photutils} package to build an effective point spread function (ePSF) using hundreds of nearby bright sources in the image (middle of Figure~\ref{fig:PS1RadProf}). 

For our analysis, we first characterized the emission around the F-type star by measuring the radial profile centered at the star. We then compared this profile with the radial profile of the ePSF scaled to the peak of the F-type star. The two radial profiles, shown in Figure~\ref{fig:PS1RadProf}, are mostly consistent with each other, particularly at the offset radius of the VLBA position (0.854\,arcsec accounting for proper motion during the PS1 observation). We also calculated the aperture magnitude of the scaled ePSF using a circular aperture with a radius of 2.56\,arcsec \citep[value obtained from the ``apRadius'' column of the PS1 Detections table;][]{Chambers2016PS1} and found $m_r = 14.0\,\mathrm{mag}$, which is consistent with the aperture magnitude of the F-type star in the PS1 catalog. Over about 8--14\,pix, the radial profile around the F-type star shows excess emission, which originates from a southwest (SW) source. We tried scaling the ePSF to match this excess emission (Figure~\ref{fig:PS1RadProf}) and found $m_r = 18.1\,\mathrm{mag}$ from aperture photometry, which is also consistent with the aperture magnitude of the SW source in the PS1 catalog. This result indicates that our ePSF is a relatively accurate representation of point sources in the PS1 image and excess optical emission from any potential counterpart of VT\,J1823-2345 is not observed at the VLBA position.

With the radial profiles of the image and the ePSF, we can roughly constrain the types of object that can remain undetected behind the observed emission by constructing and comparing radial profiles of hypothetical objects. We consider two types of objects: an unresolved point source and extended sources following a S{\'e}rsic profile. For the case of a point source, we simply shifted the radial profile of the ePSF to the VLBA location and scaled it to match the observed intensity level. We find that a point source fainter than $m_r \gtrsim 16$ will remain undetected behind the observed emission of the F-type star. 

For extended sources, we constructed 2D S{\'e}rsic profiles using \texttt{Photutils} with either an $n=1$ exponential profile typical of spiral galaxies or an $n=4$ de Vaucouleurs profile typical of elliptical galaxies. We consider an effective radius $R_e$ up to 5.0\,arcsec, which should be representative of most galaxies down to $m_r \sim 17$ \citep[][]{Shen2003}. Note that for simplicity, we do not consider ellipticity or rotation angle in the galaxy. The radial profile of the extended sources are measured after convolving the 2D S{\'e}rsic profiles with the PS1 ePSF, and three cases are shown in Figure~\ref{fig:PS1RadProf}. The magnitude of the galaxy is calculated directly from the total luminosity of the profile \citep[Equation 4 of][]{Ciotti1991}. Note that we only consider $R_e = 5.0\,\mathrm{arcsec}$ for the $n=4$ profile because the radial profile converges towards a point source for smaller $R_e$. Overall, we find that a galaxy with $m_r \gtrsim 15-16$ will be undetected at the VLBA position of J1823-2345 and brighter galaxies can be hidden for larger values of $R_e$.

\section{Interpretation as Short-Lived Peaked-Spectrum Source}\label{sec:episodicAGN}

We have considered a number of possible interpretations for VT\,J1823-2345, including an X-ray binary associated with the F-type star, a foreground or background Galactic transient, an extragalactic synchrotron transient, or a blazar. However, based on the observed properties, we argue that these scenarios are either ruled out or disfavored, and our arguments are presented in Appendix~\ref{sec:unlikelyscene}. In this Section, we focus on what we consider to be the most plausible scenario for VT\,J1823-2345 --- episodic AGN activity manifesting as a PS source. We discuss the major similarities and differences between VT\,J1823-2345 and known populations of PS sources, including the general population and other PS transients discovered in VLASS, and we briefly consider the potential implications.

\subsection{General Considerations}\label{subsec:PSgeneralconsideration}

The radio SED of VT\,J1823-2345 resembles those of PS sources, and the observed peak frequency of $\nu_{\mathrm{p,obs}} \sim6\,\mathrm{GHz}$ places VT\,J1823-2345 in the category of HFP sources. Note that blazar contamination is a known issue in variable HFP sources, but for the case of VT\,J1823-2345, arguments against blazar variability is presented in Appendix~\ref{subsec:blazarvar}. From the observed SEDs, the spectral index on the optically thick side of the radio spectrum is $\alpha_{\mathrm{thick}}\sim0.7$, which is much flatter than the expected index of 2.0 or 2.5 for synchrotron self-absorption in a homogeneous environment. This is fully consistent with PS sources because the majority of the population shows $\alpha_{\mathrm{thick}}\lesssim1.0$ \citep[e.g.,][]{Ballieux2024}, which is generally attributed to inhomogeneity in the radio-emitting shock \citep[][]{ODea1998}. The spectral index on the optically thin side is $\alpha_{\mathrm{thin}}\sim-0.9$, which can be related to the power-law index of the electron energy distribution ($N(E)\propto E^{-p}$) by $p=-2\alpha_{\mathrm{thin}}+1 = 2.8$. We also found VT\,J1823-2345 to be weakly polarized, with a linear polarization fraction of $p_{\mathrm{L}}\simeq0.73\%$ at 8.8\,GHz and $p_{\mathrm{L}}<0.25\%$ at 5.5\,GHz, which is broadly consistent with PS sources \citep[][]{ODea1998,Orienti2008}.

The spectral luminosities of PS sources at 5\,GHz (in rest frame) span many orders of magnitude, approximately over $L_{5\mathrm{GHz}} \sim 10^{22}-10^{30}\,\mathrm{W}\,\mathrm{Hz}^{-1}$ \citep[e.g.,][]{ODea1998,Slob2022,Kerrison2025}. Although we do not know the exact distance to VT\,J1823-2345, we can still infer a distance based on an assumed luminosity using the $k$-corrected spectral luminosity $L_\nu\approx4\pi d_\mathrm{L}^2S_{\nu}(1+z)^{-(1+\alpha)}$, where $\alpha$ is the spectral index. We adopt $S_\nu=70\,\mathrm{mJy}$ and use $\alpha=0.8$ to derive the rest-frame 5\,GHz spectral luminosity since the observed peak is already above 6\,GHz. If VT\,J1823-2345 is on the faint end ($L_{5\mathrm{GHz}}\lesssim10^{23}\,\mathrm{W}\,\mathrm{Hz}^{-1}$) of PS sources, it should be located at $z\lesssim0.0241$ ($d_{\mathrm{L}}\lesssim100\,\mathrm{Mpc}$). At a high luminosity of $L_{5\mathrm{GHz}}\gtrsim10^{28}\,\mathrm{W}\,\mathrm{Hz}^{-1}$, VT\,J1823-2345 would be located at $z\gtrsim3.362$ ($d_{\mathrm{L}}\gtrsim30\,\mathrm{Gpc}$). At such distances, the non-detection of optical signatures is not unexpected.

In terms of the physical structure, VT\,J1823-2345 is resolved by the VLBA and shows potentially an elongated structure corresponding to jetted outflows. If the VLBA major-axis angular size $\theta_{\mathrm{maj}}=0.4\,\mathrm{mas}$ is representative of the characteristic linear size, then we can infer the size of the VT\,J1823-2345 based on an assumed redshift and angular diameter distance $d_{\mathrm{A}}$. For a typical PS source with $L_{5\mathrm{GHz}} \sim 10^{26}\,\mathrm{W}\,\mathrm{Hz}^{-1}$, we infer a redshift $z\approx0.554$ and a projected linear size $\ell_{\mathrm{proj}}=\theta_{\mathrm{maj}}d_\mathrm{A}\approx2.7\,\mathrm{pc}$ for VT\,J1823-2345. Note that $d_{\mathrm{A}}$ reaches a maximum at around $z\approx1.59$, which implies a maximum linear size of $\ell_{\mathrm{proj,max}}\approx3.5\,\mathrm{pc}$ for VT\,J1823-2345. In other words, as an extragalactic PS source, VT\,J1823-2345 is a parsec-scale jet. 

We can also infer an average expansion speed from $\langle v \rangle \sim \ell_{\mathrm{proj}}/\Delta t = \theta_{\mathrm{maj}}d_{\mathrm{A}}(1+z)/\Delta t_{\mathrm{obs}}$. We consider the age in the observer frame to be $\Delta t_{\mathrm{obs}}\gtrsim30\,\mathrm{yr}$ since radio emission was already detected in NVSS. If we take $\Delta t_{\mathrm{obs}}=30\,\mathrm{yr}$, we find $\langle v \rangle\sim0.05c-1.0c$ for $z\sim0.05-1.64$. The apparent advancement speed should be some fraction of the speed of light for PS sources \citep[][]{ODea1998,Tschager2000,Nagai2006,OrientiDallacasa2020} and the redshift range is reasonable considering the expected luminosity. Based on these findings, a probable hypothesis is that VT\,J1823-2345 is either a low-luminosity PS source in a relatively nearby galaxy or a luminous PS source at cosmic noon.

We interpret VT\,J1823-2345 as a young, expanding jet (based on the significant brightening) and assume SSA to be the cause of the spectral turnover. Based on the spectral peak and the (major-axis) angular size, we determine the equipartition magnetic field strength following Equation~(A11) of \citet[][assuming $p=2$]{KunertBajraszewska2025}:
\begin{equation}
    B\sim18\eta_{\mathrm{eq}}^{-\frac{1}{4}}\,(1+z)^{\frac{3}{4}}\left( \frac{R}{\mathrm{pc}} \right)^{-\frac{1}{4}}\left( \frac{\nu_{\mathrm{p,obs}}}{\mathrm{GHz}} \right)^{\frac{3}{4}}\,\mathrm{mG},
\end{equation}
where $\eta_{\mathrm{eq}}$ is the equipartition ratio and $R$ is the linear size (assuming spherical symmetry). We assume equipartition $\eta_{\mathrm{eq}}=1$ for this analysis. Over a redshift range $z=0.1-5.0$ and taking $\nu_{\mathrm{p,obs}}=6\,\mathrm{GHz}$ and $\theta_{\mathrm{maj}}=0.4\,\mathrm{mas}$, we find $B_{\mathrm{eq}}\sim80-210\,\mathrm{mG}$. This field strength is broadly consistent with PS sources \citep[e.g.,][]{ODea1998,Orienti2008eq}, although we emphasize that there are significant uncertainties in this derivation because we assumed equipartition and derived the size based on a Gaussian fit at 15\,GHz. 

Assuming this estimate of magnetic field strength is roughly accurate, we consider the spectral age and the break frequency for VT\,J1823-2345 following Equation~(2) of \citet[][]{Murgia1999}:
\begin{equation}
    t_{\mathrm{syn}} = 1610\frac{B^{1/2}}{B^2+B_{\mathrm{CMB}}^2}\frac{1}{[\nu_{\mathrm{br,obs}}(1+z)]^{1/2}}\,\mathrm{Myr},
\end{equation}
where $t_{\mathrm{syn}}$ is the spectral age, $B$ is the magnetic field strength in $\mu$G, $B_{\mathrm{CMB}}=3.25(1+z)^2\mathrm{\mu\mathrm{G}}$, and $\nu_{\mathrm{br,obs}}$ is the break frequency in GHz in the observer frame. If we assume $B=B_{\mathrm{eq}}$ for VT\,J1823-2345 and the age (in the observer frame) to be on the order of $t_{\mathrm{syn,obs}}\sim10-100\,\mathrm{yr}$, then we find $\nu_{\mathrm{br,obs}}\sim0.8-80\,\mathrm{GHz}$ at $z=1.0$. A smaller age corresponds to a higher break frequency, and the values change slightly for varying redshifts ($\nu_{\mathrm{br,obs}}$ can become an order of magnitude smaller at very high $z\gtrsim9$). If this analysis is roughly accurate, then the lack of an observed spectral break below 15\,GHz (Figure~\ref{fig:RadioSEDs}) implies that the spectral age of VT\,J1823-2345 is probably less than a couple of decades, and we would expect $\nu_{\mathrm{br,obs}}$ to be around sub-centimeter wavelengths. In this case, the constraint on the spectral age supports the interpretation that VT\,J1823-2345 is a young PS source.

With the magnetic field strength, we can also make a rough order-of-magnitude estimate of the ambient density. We assume the internal energy density $u_{\mathrm{int}}$ is related to the magnetic energy density $u_{\mathrm{B}}$ by $u_{\mathrm{int}}=u_B/\epsilon_B=B^2/(8\pi\epsilon_B)$. Based on the assumption of a strong shock \citep[following the analysis in Section~3.2 of][]{Ho2019} and adopting an adiabatic index of $5/3$, we estimate the mass density of the surrounding medium to be $\rho_{\mathrm{ext}} = B^2/(9\pi\epsilon_Bv^{2})$, where $v$ is the shock expansion speed. Assuming the surrounding medium is composed of fully ionized hydrogen, then $\rho_{\mathrm{ext}}=\mu m_pn_{\mathrm{ext}}$, where $\mu=1$, $m_p$ is the proton mass, and $n_{\mathrm{ext}}$ is the ambient number density, we find:
\begin{equation}
    n_{\mathrm{ext}} \simeq 2.4\times 10^{2}
    \left(\frac{\epsilon_B}{0.1}\right)^{-1}
    \left(\frac{B}{100\,\mathrm{mG}}\right)^2
    \left(\frac{v}{0.1c}\right)^{-2}\,\mathrm{cm^{-3}}.
\end{equation}
Based on this, the external density can be quite high ($n_{\mathrm{ext}}\gtrsim10^3\,\mathrm{cm^{-3}}$) for a high magnetic field strength ($B\gtrsim100\,\mathrm{mG}$) and a relatively low expansion speed ($v\lesssim0.1c$). A high density would be inconsistent with the interstellar medium and would perhaps favor an AGN \citep[e.g., the narrow-line region;][]{Kakkad2018}, although it is difficult to rule out parameters that would yield $n_{\mathrm{ext}}\sim1\,\mathrm{cm^{-3}}$ with our current observations. We emphasize again that this analysis should be treated, at best, as an order-of-magnitude estimate because of the simplifying assumptions (e.g., equipartition) and the neglect of relativistic and geometric effects. Nonetheless, we present this analysis to demonstrate that there is likely a reasonable parameter space where the ambient density is consistent with the nuclear environment of AGN, which is expected for PS sources. 

In summary, based on the radio SEDs and the VLBA observations, we can reasonably interpret VT\,J1823-2345 as a young PS source with a parsec-scale jet that is likely decades old. The distance to VT\,J1823-2345 is uncertain, but it could either be a low-luminosity PS source in a relatively nearby galaxy or a luminous PS source at the cosmic noon.

\subsection{Time Evolution and Comparison with Correlations}\label{subsec:timeevolcorrelation}

Comparing VT\,J1823-2345 to the general population of PS sources, the biggest discrepancy is perhaps the radio variability. Generally, at centimeter wavelengths, PS sources exhibit variability within a couple of percent, up to about 30\% over years \citep[e.g.,][]{ODea1998,Edwards2004,OrientiDallacasa2020,Ross2021,Kerrison2025}. The most extreme variability, particularly in the spectral shape, is usually attributed to blazar variability, which we do not consider here (see Section~\ref{subsec:blazarvar}). Moderate variability without significant changes in the spectral shape is generally attributed to interstellar scintillation or intrinsic evolution of the expanding jet \citep[][]{OrientiDallacasa2020,Ross2021}. In contrast, VT\,J1823-2345 has brightened by about a factor of 4.3 at 1.4\,GHz between 1996 and 2024 and by a factor of 1.8 at 3.0\,GHz between 2019 and 2023 before fading (Figure~\ref{fig:RadioLCs}). The radio SED also shifted noticeably between 2023 and 2024. Therefore, VT\,J1823-2345 would constitute an extreme outlier in the population of PS sources in terms of variability.

However, this discrepancy may simply be due to the age difference between VT\,J1823-2345 and most PS sources. A significant change in the spectral peak can be interpreted as an expanding jet if VT\,J1823-2345 is only a few decades old, which is probable considering the observed radio properties (see Section~\ref{subsec:PSgeneralconsideration}). Conversely, the majority of PS sources are likely orders of magnitude older \citep[][]{ODeaSaikia2021} and thus should not exhibit significant variability over a timescale of years. 

To illustrate this point here, we estimate the kinematic age of VT\,J1823-2345 from the change in flux density in the optically thick regime, following Section~3.3 of \citet[][]{OrientiDallacasa2020}:
\begin{equation}\label{eq:Sratio_brighten}
    \frac{S_1}{S_0} = \left( \frac{t_0+\Delta t}{t_0} \right)^{3},
\end{equation}
where $S_0$ is the flux density at an age of $t_0$, and $S_1$ is the value at a later time of $t_0 + \Delta t$. We can also examine the expected change in peak frequency, following \citet{vanderLaan1966} to incorporate the electron energy distribution:
\begin{equation}\label{eq:nupratiodeltat}
    \frac{\nu_{\mathrm{p,1}}}{\nu_{\mathrm{p,0}}} = \left( \frac{t_0}{t_0+\Delta t} \right)^{\frac{4p+6}{p+4}}.
\end{equation}
Again, the derived ages should be treated as order-of-magnitude estimates due to the simplifying assumptions \citep[as discussed by][]{OrientiDallacasa2020}.

For the flux density ratio, we use the two 1.4\,GHz measurements from the 1996 NVSS observation and the 2024 VLA observation, which are separated by about $\Delta t_{\mathrm{obs}}\simeq28\,\mathrm{yr}$. From this flux density ratio, we find a kinematic age of $t_{0,\mathrm{obs}}\simeq44\,\mathrm{yr}$ in 1996 and $t_{0,\mathrm{obs}}+\Delta t_{\mathrm{obs}}\simeq72\,\mathrm{yr}$ in 2024, both in the observer frame. 

\begin{figure*}
    \centering
    \includegraphics[width=0.49\textwidth]{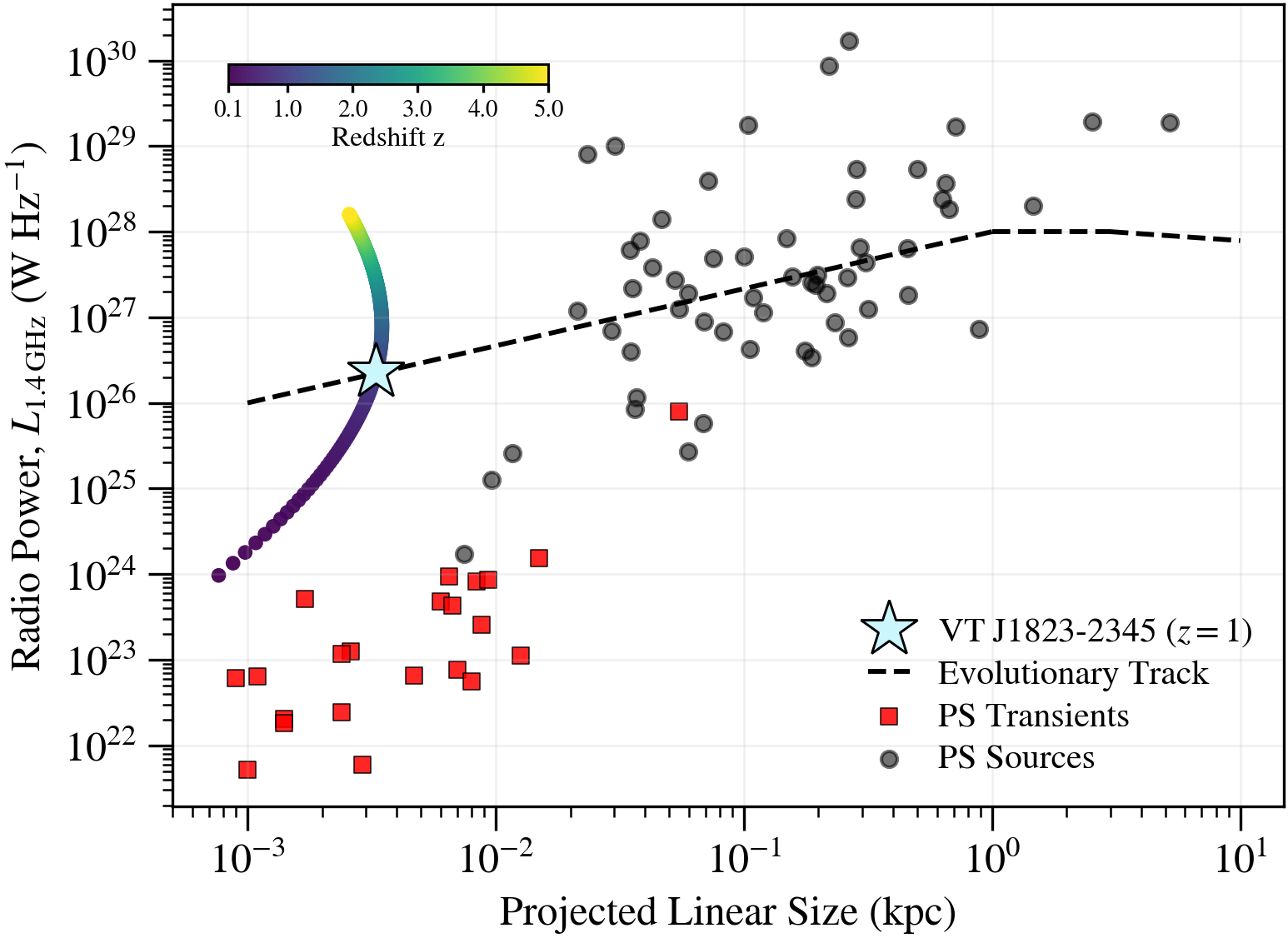}
    \includegraphics[width=0.49\textwidth]{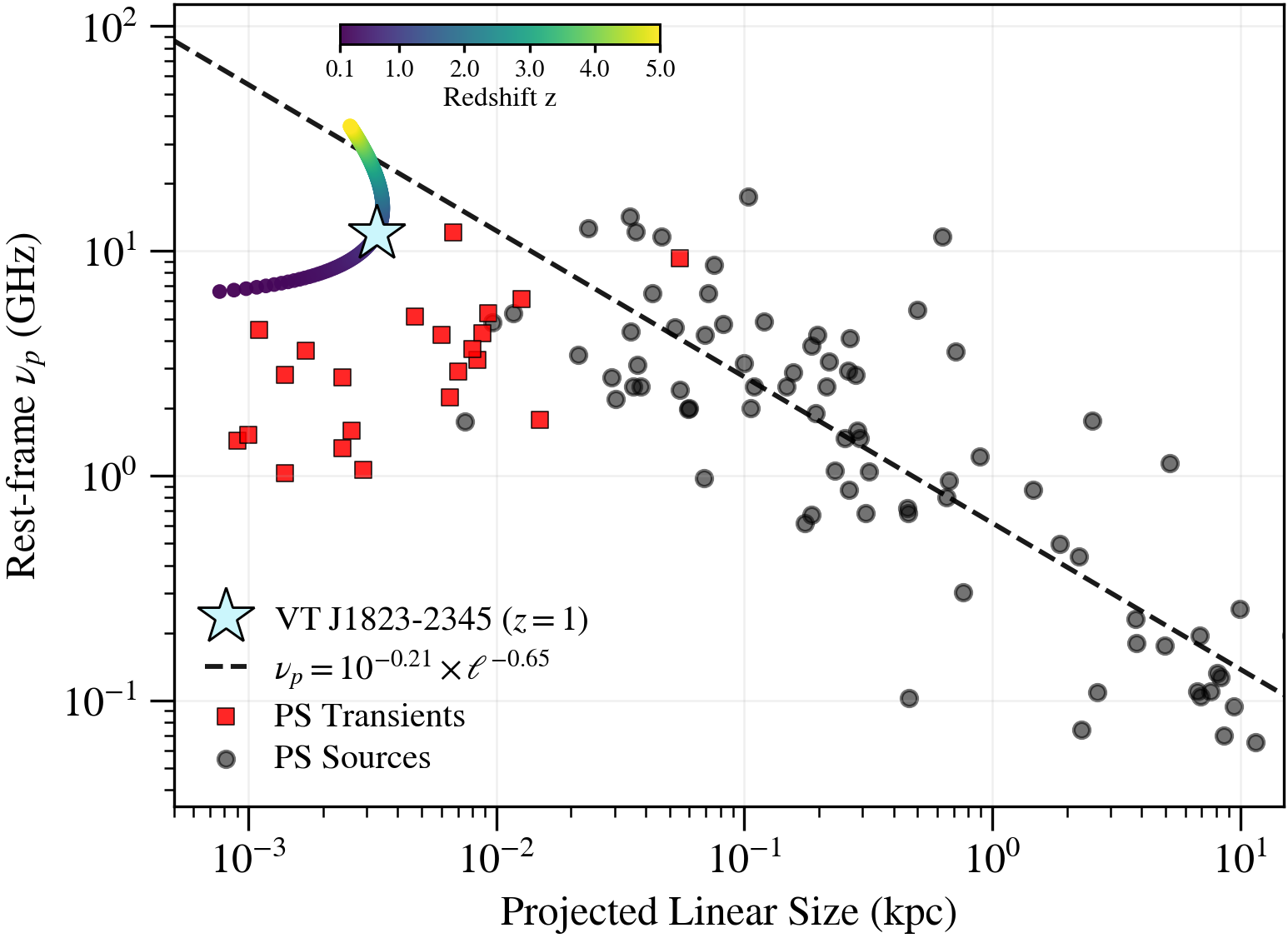}
    \caption{Left panel: radio power (1.4\,GHz) vs. projected linear size for PS sources. Right panel: the rest-frame peak frequency vs. projected linear size for PS sources. In both panels, the location of VT\,J1823-2345 is marked by a blue star assuming $z=1$ along with points derived for $z=0.1-5.0$ over a range of colors (see color map at the top). For comparison, PS sources from \citet{Keim2019} are shown as gray circles, compiled from a number of references \citep[][]{Fanti1990,deVries1997,ODeaBaum1997,Snellen1998,Stanghellini1998,Snellen2000}. In addition, a number of probable PS sources identified as VLASS transients from \citet{KunertBajraszewska2025} are shown as red squares. On the left, an example evolutionary track for high-power radio sources is shown as a dashed line, taken from \citet{An&Baan2012}. On the right, a best-fit line for PS sources is shown as dashed line, taken from \citet{ODeaBaum1997}.}
    \label{fig:pDnupD}
\end{figure*}

Then, with the inferred age, we can estimate the expected change in peak frequency between the ATCA and VLA observations. If we take $t_{0,\mathrm{obs}}+\Delta t_{\mathrm{obs}}\simeq72\,\mathrm{yr}$ for the VLA observation in 2024 and $t_{0,\mathrm{obs}}\simeq70.7\,\mathrm{yr}$ and $\nu_{\mathrm{p,0}}=6.65\,\mathrm{GHz}$ for the ATCA observation in 2023 (Table~\ref{tab:radioproperties}), then we expect $\nu_{\mathrm{p,1}}\sim6.4\,\mathrm{GHz}$ to be the peak frequency measured by VLA in 2024. This value is very much within the uncertainty of the observed $\nu_{\mathrm{p}}$ in 2024. Therefore, these estimates demonstrate that a decades-old jet can explain the observed variability of VT\,J1823-2345.

Based on the observed properties, we can also consider VT\,J1823-2345 under the hypothesized evolutionary path of young PS sources and radio-loud AGN. In particular, at the early evolutionary stage, as the jet undergoes adiabatic expansion, the increase in source size should correlate with a decrease in peak/turnover frequency and an increase in radio power \citep[][]{Snellen2000,KunertBajraszewska2010,An&Baan2012}. These correlations have been examined empirically for PS sources and radio-loud AGN \citep[e.g.,][]{ODea1998,KunertBajraszewska2010,Keim2019,KunertBajraszewska2025}. To investigate how VT\,J1823-2345 fits within this understanding, we compare its properties to these correlations.

In Figure~\ref{fig:pDnupD}, we show the correlation between the 1.4\,GHz radio power $L_{1.4\mathrm{GHz}}$ and the projected linear size $\ell_{\mathrm{proj}}$ (left panel) as well as the correlation between the rest-frame peak frequency $\nu_\mathrm{p}$ and the projected linear size $\ell_{\mathrm{proj}}$ (right panel). VT\,J1823-2345 is shown as a star for $z=1$, along with points derived for $z=0.1-5.0$ over a range of colors. Other PS sources from the literature are also shown for reference \citep[][]{Keim2019}. In addition, we show transient sources identified in VLASS by \citet{KunertBajraszewska2025} that exhibit peaked spectra and compact sizes, which may be associated with PS sources.

On the $L_{1.4\mathrm{GHz}}$ vs. $\ell_{\mathrm{proj}}$ diagram (left panel of Figure~\ref{fig:pDnupD}), we show an example evolutionary path for high-power radio sources \citep[][]{An&Baan2012}, which follows the radio power as the jet expands. Over this path, the youngest jets are expected to brighten while undergoing adiabatic expansion, and they will begin to fade as synchrotron radiation losses start to dominate at larger physical scales. Eventually, the jet will evolve into large-scale radio galaxies. VT\,J1823-2345 is broadly consistent with this evolutionary path, but at a parsec scale compared to $\ell_{\mathrm{proj}} \gtrsim 10-1000\,\mathrm{pc}$ of typical PS sources, which places it in the earliest brightening phase. This is consistent with the observed radio light curves, which continue to brighten at low frequencies. Compared to the transient PS sources identified by \citet{KunertBajraszewska2025}, VT\,J1823-2345 is orders of magnitude more luminous unless it is at very low redshift ($z\lesssim0.1$). \citet{KunertBajraszewska2025} speculated that these transient PS sources may be a low-luminosity population that evolves separately. In contrast, VT\,J1823-2345 is perhaps more closely related to the high-power radio sources than these other transient PS sources, at least on this diagram. 

On the $\nu_{\mathrm{p}}$ vs. $\ell_{\mathrm{proj}}$ diagram (right panel of Figure~\ref{fig:pDnupD}), we show a best-fit line that represents the empirical trajectory of PS sources \citep[][]{ODeaBaum1997}, i.e., the decline of the peak frequency as the jet expands. On this diagram, VT\,J1823-2345 resides near the top left and represents an extremely compact HFP source, consistent with a young jet. However, compared to the extrapolated empirical trajectory, VT\,J1823-2345 appears to have a lower $\nu_\mathrm{p}$ unless it is at a high redshift ($z\gtrsim4$). Or, from a different perspective, VT\,J1823-2345 appears to be smaller in size compared to PS sources with the same $\nu_\mathrm{p}$. Here, the transient PS sources are also below the empirical trajectory, and VT\,J1823-2345 appears close to this population. We do note that the scattering around the correlation is large, so the comparisons here are mostly qualitative.

As seen in these correlations, the expectation is that VT\,J1823-2345 (as a young jet) should continue to brighten at low frequencies until it reaches about a kiloparsec in scale, and the peak frequency will continue to drop as the emission becomes optically thin. However, based on the 3.0\,GHz light curve (observer frame), the gigahertz emission of VT\,J1823-2345 is already beginning to fade. This raises the question of whether VT\,J1823-2345 will follow the expected trajectory, especially considering that its peak frequency may be lower than expected. 

To illustrate this, we again consider the change in peak frequency using Equation~\ref{eq:nupratiodeltat}. Specifically, we can estimate the age when the peak frequency crosses 1.4\,GHz in the rest frame, $t(\nu_{\mathrm{p}}=1.4\,\mathrm{GHz})$, assuming this is when $L_{\mathrm{1.4GHz}}$ begins to drop as the emission becomes optically thin. For this calculation, we adopt $\nu_{\mathrm{p,0}}=6.6(1+z)\,\mathrm{GHz}$, $\nu_{\mathrm{p,1}}=1.4\,\mathrm{GHz}$, $p=2$, and $t_0=72/(1+z)\,\mathrm{yr}$ (based on the previous analysis of variability) to estimate $t_0 + \Delta t$ in the rest frame. We find $t(\nu_{\mathrm{p}}=1.4\,\mathrm{GHz}) \sim 28-99\,\mathrm{yr}$ for VT\,J1823-2345, assuming $z\sim 5.0-0.1$. According to these estimates, even if VT\,J1823-2345 is expanding relativistically, it will reach at most a few dozen parsecs before fading at 1.4\,GHz. This analysis perhaps suggests that rather than following the trajectory of high-power radio sources, VT\,J1823-2345 is a ``short-lived'' PS sources that will not grow into a large-scale radio galaxy. 

This interpretation is in line with the hypothesis that some PS sources are episodic transient sources that dissipate before reaching large physical scales \citep[][]{ODeaBaum1997,ReynoldsBegelman1997,KunertBajraszewska2010}. \citet{KunertBajraszewska2025} also considered this explanation for the transient PS sources and argued that an outburst duration of decades to centuries is plausible for black holes with relatively low masses if the intermittent activity is caused by radiation pressure instability in the accretion disk \citep[][]{Czerny2009}. VT\,J1823-2345 is perhaps similar to these transient PS sources and may instead evolve into a low-luminosity radio source. The launching of low-luminosity jet has also been observed in changing-look AGN \citep[][]{Meyer2025,Laha2025}, which resemble PS sources and may be what causes episodic radio AGN \citep[although see][]{Birmingham2025}. We also note that relativistic or jetted TDEs are another known population of transients that exhibit luminous synchrotron radiation with peaked spectra evolving over decadal timescales \citep[e.g.,][]{Berger2012Sw1644,Zauderer2013SwJ1644,Eftekhari2018Sw1644,Andreoni2022AT2022cmc,Rhodes2023AT2022cmc,Eftekhari2024AT2022cmc,Cendes2021J1644}. Rather than AGN activity, jetted TDEs may also contribute to the population of episodic PS sources \citep[][]{Readhead2024}.

Finally, we briefly consider the fading of VT\,J1823-2345 at 3.0\,GHz in the observer frame following the predicted evolution \citep[][]{vanderLaan1966}: 
\begin{equation}
    \frac{S_1}{S_0} = \left( \frac{t_0+\Delta t}{t_0} \right)^{-2p}.
\end{equation}
Here, we assume that the emission at 3.0\,GHz has peaked and has become optically thin during the ATCA observation in 2023. In the observer frame, taking $S_0 = 58.8\,\mathrm{mJy}$, $t_0 = 70.7\,\mathrm{yr}$ (based on analysis using Equation~\ref{eq:Sratio_brighten}), and $p=2$, we find that the 3\,GHz will reach $S_1\simeq0.1\,\mathrm{mJy}$ in about 277\,yr, or in the year 2300, below the VLASS detection threshold. In other words, if VLASS continued to monitor the 3\,GHz sky over centuries (just as an example), VT\,J1823-2345 would appear as a ``reverse'' transient source that will fade and eventually disappear. Therefore, from an observational perspective, VT\,J1823-2345 will manifest as a short-lived radio source compared to ordinary radio AGN on the sky. 

In summary, we find that the radio variability of VT\,J1823-2345 is consistent with the expansion of a young, decades-old jet. Comparisons with known correlations place VT\,J1823-2345 at the earliest evolutionary phase of PS sources and radio galaxies. However, based on the predicted evolution for adiabatic expansion, we find that VT\,J1823-2345 may begin to fade at centimeter wavelengths in the rest frame well before it reaches the typical sizes of PS sources. In the observer frame, VT\,J1823-2345 would manifest as a reverse transient source in any VLASS-like survey and disappear within a few centuries. Therefore, it is probable that VT\,J1823-2345 represents a case of short-lived PS source related to episodic AGN activity.

\subsection{Comparisons with Literature VLASS Nuclear Transients}

A number of studies have identified nuclear transients in VLASS \citep[e.g.,][]{Nyland2020,Zhang2022,Chen2024InvertedTransient,KunertBajraszewska2025}, primarily by comparing VLASS to NVSS and the Faint Images of the Radio Sky at Twenty-Centimeters \citep[FIRST;][]{Becker1995FIRST,Helfand2015FIRST14Dec17} survey. FIRST was another survey conducted with the VLA at 1.4\,GHz over 1993--2004 and 2009--2011, which, compared to NVSS, covered less area of the sky but with a higher resolution (5\,arcsec for FIRST vs. 45\,arcsec for NVSS) and a better sensitivity (0.15\,mJy\,beam$^{-1}$ of FIRST vs. 0.45\,mJy\,beam$^{-1}$ of NVSS). In general, these studies considered VLASS transients to be radio sources detected in VLASS but not detected in FIRST or NVSS. Nuclear transients were identified by associating the radio transients to galaxies and quasars. The origin of these VLASS nuclear transients was thought to be related to AGN activity or TDEs, and PS sources were often considered in the interpretations. 

Here, we briefly compare VT\,J1823-2345 to VLASS nuclear transients reported in the literature. In particular, we compare the light curves since they were all discovered in the same survey over similar time frames. For this comparison, we compiled 69 VLASS nuclear transients reported by \citet{Nyland2020}, \citet{Zhang2022}, \citet{KunertBajraszewska2025}, and \citet{Chen2024InvertedTransient}. Note that the selections made by these studies were not homogeneous and were generally limited in scope: \citet{Nyland2020} specifically selected quasars in the FIRST coverage, \citet{Zhang2022} selected SDSS-classified galaxies at $z<0.3$ in the FIRST coverage, \citet{KunertBajraszewska2025} selected bright VLASS transients associated with optical hosts, and \citet{Chen2024InvertedTransient} selected bright VLASS transients based on radio spectral index and not host association. Here, we only collected transients from \citet{Chen2024InvertedTransient} that had galaxy associations. 

For these transients, we collected the full VLASS light curves from the VLASS QL point source catalog, which is a high-quality VLASS catalog that will be presented in a forthcoming paper (D. Dong et al., in prep.). The details will be presented in the paper, but in summary, this point source catalog was constructed using a customized pipeline that identifies an initial source list using the Python Blob Detector and Source Finder \citep[PyBDSF;][]{MohanRafferty2015PyBDSF} and employs a number of algorithms for quality assurance (e.g., artifact rejection). Only point sources (consistent with the restoring beam) are extracted and bootstrapping techniques are used to characterize faint sources near the detection threshold. Flux density measurements from all epochs (epochs 1, 2, 3, and 4.1) have been compiled for each source, including both detections and non-detections. The main purpose of this catalog is to identify variables and transients within VLASS, so tests and corrections have been performed to ensure that flux density extraction is consistent and robust over all epochs. 

\begin{figure}
    \centering
    \includegraphics[width=0.45\textwidth]{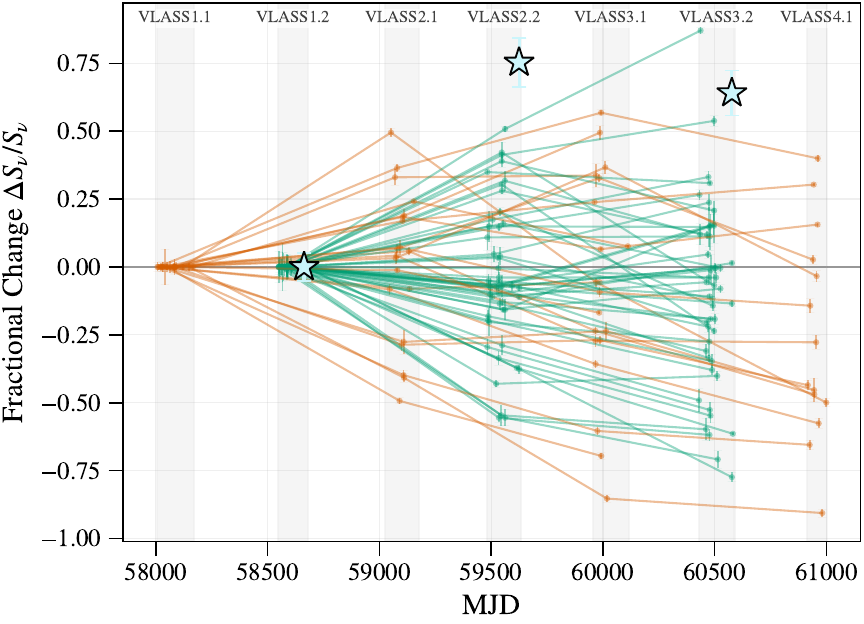}
    \caption{The fractional change in the 3\,GHz VLASS light curve for VLASS nuclear transients collected from the literature \citep[references:][]{Nyland2020,Zhang2022,KunertBajraszewska2025,Chen2024InvertedTransient}, shown as circles connected by lines. The date range for each VLASS half-epoch is shaded. For each light curve, the fractional change is derived relative to the flux density at epoch 1. Different colors are used to distinguish transients observed during the first and second half-epochs. The changes in VT\,J1823-2345 are marked by stars.}
    \label{fig:LitNucVT}
\end{figure}

Using the VLASS QL point source catalog, we collected 3.0\,GHz VLASS light curves for 68 VLASS nuclear transients reported in the literature (one was not found in the catalog, likely due to quality cuts). In Figure~\ref{fig:LitNucVT}, we show the fractional change in the 3\,GHz flux density $\Delta S_\nu/S_\nu$ over the MJD of the observing date for the nuclear transients. The fractional change for each light curve is calculated relative to the flux density at the first epoch, such that $\Delta S_\nu/S_\nu=0$ at epoch 1. Note that VLASS epochs are divided into half-epochs (e.g., epoch 1 is made up of epoch 1.1 and 1.2), and observations of the same locations are separated by one full epoch. The changes in VT\,J1823-2345 over VLASS are marked by stars. 

Over the full survey, the majority of VLASS nuclear transients reported in the literature vary within $\lesssim$\,30\%--50\% over $\sim$\,2000--3000\,days, with a small fraction showing $|\Delta S_\nu/S_\nu| \gtrsim 0.5$. Compared to most of these nuclear transients, VT\,J1823-2345 appears to exhibit a greater degree of brightening in VLASS (with $\Delta S_\nu/S_\nu \sim 0.75$ between epoch 1 and 2), but it does not appear to be an extreme outlier. We note that the VLASS nuclear transients typically showed a 3.0\,GHz flux density of at least a few mJy, up to tens of mJy. At this brightness, the typical rms of the VLASS QL images ($\sim$\,0.14\,$\mu$Jy) contributes very little (at most a few percent) to the random variability of VLASS sources. Therefore, much of the fractional changes shown in Figure~\ref{fig:LitNucVT} should represent genuine source variability. It is also worth noting that there are more occasions of fading than brightening. Between epochs, we find 39 occasions of brightening and 72 occasions of fading above the $3\sigma$ level, where $\sigma$ is the combined flux density uncertainty. 

Additional properties were also reported for these transients, but available constraints differ between studies. Transients reported by \citet{Chen2024InvertedTransient} and \citet{Zhang2022} generally lack precise constraints on the radio spectra and projected size. \citet{Nyland2020} reported radio SEDs for their transients but had no high-resolution observations to constrain sizes. Only \citet{KunertBajraszewska2025} reported both radio SEDs and high-resolution VLBA observations for their sample. Redshifts were generally available in the collected sample, with the majority of quasars reported at high redshifts \citep[$z\gtrsim1$;][]{Nyland2020} and the majority of nearby galaxies ($z\lesssim0.3$) not hosting quasars \citep[][]{Zhang2022,KunertBajraszewska2025}. Among the reported radio SEDs, most VLASS nuclear transients exhibit peaked radio spectra. The majority of quasars peak at high frequencies with $\nu_{\mathrm{p,obs}}\sim5-22\,\mathrm{GHz}$ \citep[or $\nu_{\mathrm{p}}\sim12-56\,\mathrm{GHz}$ in the rest frame;][]{Nyland2020}. On the other hand, the majority of galaxies peak at $\nu_{\mathrm{p}}\lesssim1-5\,\mathrm{GHz}$ \citep[][]{KunertBajraszewska2025}.

Physically, if we assume that the radio variability of these nuclear transients is associated with young expanding jets, then our analysis in Section~\ref{subsec:timeevolcorrelation} also applies to these transients. In this case, the prevalence of fading transients indicates that for many of these transients, the 3.0\,GHz emission (in the observer frame) has become optically thin, with the radio spectrum peaking at $\nu_{\mathrm{p,obs}}\lesssim3.0\,\mathrm{GHz}$. This is consistent with the reported peak frequencies, which are generally near or below 3.0\,GHz for the low-redshift galaxies that dominate the sample. This also explains why there are fewer cases of brightening similar to VT\,J1823-2345, because for many, significant brightening at 3.0\,GHz occurred prior to the VLASS observations. According to our previous analysis (Section~\ref{subsec:timeevolcorrelation}), the level of variability seen in these VLASS nuclear transients implies a kinematic age of around decades. Assuming an expansion speed on the order of $\langle v \rangle \sim 0.1c$, the expected size should be on the order of $\ell\lesssim10\,\mathrm{pc}$, which was found to be the case for most transients reported by \citet{KunertBajraszewska2025}. 

For the quasars, the high-frequency peak, the high spectral luminosity, and the (speculated) compact sizes would be consistent with the early phase of high-power radio sources. If they continue to brighten, they would follow the predicted evolution of PS sources that grow into large-scale radio galaxies. However, for most of the low-redshift galaxies, the combination of $\nu_{\mathrm{p}}\lesssim3\,\mathrm{GHz}$ and $\ell\lesssim10\,\mathrm{pc}$ is located well below the extrapolated empirical trajectory \citep[as shown for those reported by][see Figure~\ref{fig:pDnupD}]{KunertBajraszewska2025}. In addition, if the low-redshift galaxies are truly fading at the predicted rate for adiabatic expansion, then like VT\,J1823-2345, they will appear as transients at centimeter wavelengths over a timescale of decades to centuries (Section~\ref{subsec:timeevolcorrelation}). Therefore, much like VT\,J1823-2345, we can speculate based on variability that many of the reported low-redshift VLASS nuclear transients are perhaps also short-lived transient PS sources. In this case, VT\,J1823-2345 is a part of this same group of transients, but only appears different in VLASS because it was caught during the brightening phase when the (observer frame) 3.0\,GHz emission was still brightening and optically thick.

We note that for this particular VLASS sample of nuclear transients, the prevalence of fading transients is likely related to intrinsic properties and selection bias. In particular, their selection is biased towards low-redshift galaxies with extreme radio variability within a timescale of about two decades. This could lead to the preferential selection of low-power PS sources that evolve on such short timescales \citep[as discussed by][]{KunertBajraszewska2025}. The selection of VT\,J1823-2345 is slightly different because it was not selected based on optical association and a non-detection in NVSS was not required. As a result, it appears closer to the slow-evolving high-power PS sources (Figure~\ref{fig:pDnupD}), which may explain the apparent difference between VT\,J1823-2345 and many of the reported VLASS nuclear transients.

In summary, we find that VT\,J1823-2345 showed a greater degree of brightening compared to most VLASS nuclear transients reported in the literature, but it does not appear to be an extraordinary outlier. Considering the variability and spectral peak, we speculate that many of the reported VLASS nuclear transients (associated with low-redshift galaxies) may also be short-lived transient PS sources, and VT\,J1823-2345 would belong in this same group of transients. We emphasize that at this stage, our discussion here is very much speculative due to lack of constraints on the spectral peak, spectral evolution, physical size, and expansion speed for many of the reported transients. However, if our speculation is roughly accurate, then radio surveys are beginning to reveal populations of transient PS sources as the temporal baselines of these surveys extend over decades. Since earlier searches for radio nuclear transients were generally limited in scope, we suspect that there exists a much larger population of transient PS sources that may be discoverable in surveys such as VLASS when the full scope of the survey is considered. In this case, multi-epoch radio continuum and VLBI observations will be crucial for characterizing these transients and expanding our understanding of jet launching and early jet evolution in AGN.

\section{Summary and Conclusion}\label{sec:conclusion}

In this study, we presented our observations and analyses on the transient VT\,J1823-2345 discovered in VLASS and VCSS. Our findings through archival and follow-up observations are summarized below.
\begin{itemize}
    \item VT\,J1823-2345 was discovered to have brightened significantly (by a factor of a few) between NVSS, VLASS, and RACS-mid between 1996 and 2021. A non-detection in VCSS made it a candidate inverted-spectrum transient with $\alpha\gtrsim0.35$ between 340\,MHz and 3.0\,GHz.
    \item VT\,J1823-2345 exhibited peaked radio SEDs in our follow-up ATCA and VLA observations in 2023 and 2024, peaking at around $\nu_{\mathrm{p,obs}}\sim6.65\,\mathrm{GHz}$. The peak flux density is decreasing over time. The spectral indices were about $\alpha_{\mathrm{thick}}\sim0.7$ and $\alpha_{\mathrm{thin}}\sim-1.0$ (see Figure~\ref{fig:RadioSEDs} and Table~\ref{tab:radioproperties}). 
    \item VT\,J1823-2345 is weakly polarized, with a linear polarization fraction of $p_{\mathrm{L}}\simeq0.73\%$ at 8.8\,GHz and $p_{\mathrm{L}}<0.25\%$ at 5.5\,GHz. The decrease in polarization fraction over wavelength implies intrinsic depolarization. An extreme $\mathrm{RM_{obs}}\sim12,000\,\mathrm{rad}\,\mathrm{m^{-2}}$ in the observer frame was also found (see Figure~\ref{fig:RMsynth}).
    \item In the observer frame, the 3\,GHz light curve of VT\,J1823-2345 has peaked around 2023 and has been fading since then, with an inferred time above half-maximum of about 10 years. The 1.4\,GHz and 887.5\,GHz continued to brighten.
    \item Our VLBA observations in 2024 revealed a resolved Gaussian size at 15\,GHz of $\theta_{\mathrm{maj}}\times\theta_{\mathrm{min}}=0.40\,\mathrm{mas}\times0.08\,\mathrm{mas}$, which yielded a brightness temperature of $T_\mathrm{B}=5.7\times10^9(1+z)\,\mathrm{K}$. An additional VLBA snapshot in 2026 also showed a resolved source with comparable sizes at 4.3\,GHz and 7.6\,GHz.
    \item X-ray emission was detected at the location of VT\,J1823-2345, with an approximate power-law photon index of $\Gamma_{\mathrm{X}}\sim1.5\pm0.3$. No significant X-ray variability was found between 2011 and 2023.
    \item On the sky, VT\,J1823-2345 appears coincident with an F-type star at 1.4\,kpc, near the projected center of the SNR G8.7-5.0 (Figure~\ref{fig:radoptimg}). However, based on the VLBA coordinates, VT\,J1823-2345 is separated from the F-type star by 0.914\,arcsec. No optical or infrared flares were found in ATLAS, PS1, or WISE. On the PS1 image, the F-type star can hide a point source with $m_{\mathrm{AB}}\gtrsim16.5$ in the r-band at the location of VT\,J1823-2345 (Figure~\ref{fig:PS1RadProf}). 
\end{itemize}

We argued that VT\,J1823-2345 is best interpreted as a young peaked-spectrum source. Based on its angular size, lack of a spectral break, and temporal evolution, we found that VT\,J1823-2345 is consistent with a decades-old jet. Compared to the general population of PS sources, VT\,J1823-2345 is at the earliest evolutionary stage of radio AGN, but it may disappear on the GHz radio sky over centuries, well before it grows into a large-scale radio galaxy. Therefore, we argued that VT\,J1823-2345 is a short-lived transient PS source possibly related to episodic AGN activity or jetted TDEs.

Through comparing VT\,J1823-2345 to other nuclear transients discovered in VLASS, we speculated that it belongs to an emerging class of transient PS sources, which are now discoverable in radio surveys that cover temporal baselines of years and decades. These transients offer direct insights into the youngest jets from supermassive black holes, the early evolution of radio sources, and the activity cycles of AGN. Multi-epoch radio continuum and VLBI observations will be crucial for characterizing these radio jets and establishing an important discovery space in time-domain AGN research for upcoming facilities such as DSA-2000 \citep[][]{Hallinan2019DSA2000}, the SKA \citep[][]{Braun2019SKA}, and the ngVLA \citep[][]{Murphy2018ngVLA}. 

\begin{acknowledgments}

We thank the anonymous reviewer for comments and feedback that improved the content and organization of this manuscript. We thank Jamie Stevens for scheduling the ATCA observation on green time and the NRAO staff for scheduling the VLA and VLBA observations. We also thank Gregory Sivakoff for help on the VLA proposal. Y.C. would like to thank Mattias Lazda and Ayush Pandhi for suggestions on the radio data reduction and analysis. Y.C. acknowledges support from the Natural Sciences and Engineering Research Council of Canada (NSERC) Canada Graduate Scholarships – Doctoral Program. The Dunlap Institute is funded through an endowment established by the David Dunlap family and the University of Toronto. Parts of this research were supported by the Australian Research Council Discovery Early Career Researcher Award (DECRA) through project number DE230101069.

AI tools were used for generating or refining code used for the analyses presented in this manuscript. Certain figures were also produced by code generated or refined by AI tools. AI tools have not been used to directly generate any text or tables in this manuscript. The GAI tools used were ChatGPT and Claude.

The VLA is operated by the National Radio Astronomy Observatory (NRAO). The NRAO is a facility of the U.S. National Science Foundation operated under cooperative agreement by Associated Universities, Inc. This work made use of the Swinburne University of Technology software correlator, developed as part of the Australian Major National Research Facilities Programme and operated under licence. Basic research in radio astronomy at the U.S. Naval Research Laboratory (NRL) is supported by 6.1 Base funding. Construction and installation of VLITE was supported by the NRL Sustainment Restoration and Maintenance fund. CIRADA is funded by a grant from the Canada Foundation for Innovation 2017 Innovation Fund (Project 35999), as well as by the Provinces of Ontario, British Columbia, Alberta, Manitoba and Quebec. This research has made use of the VizieR catalogue access tool, CDS, Strasbourg, France \citep{10.26093/cds/vizier}. The original description of the VizieR service was published in \citet{vizier2000}.

The Australia Telescope Compact Array is part of the Australia Telescope National Facility (grid.421683.a) which is funded by the Australian Government for operation as a National Facility managed by CSIRO. We acknowledge the Gomeroi people as the traditional owners of the Observatory site. This scientific work uses data obtained from Inyarrimanha Ilgari Bundara / the Murchison Radio-astronomy Observatory. We acknowledge the Wajarri Yamaji People as the Traditional Owners and Native Title Holders of the Observatory site. CSIRO’s ASKAP radio telescope is part of the Australia Telescope National Facility (\url{https://ror.org/05qajvd42}). Operation of ASKAP is funded by the Australian Government with support from the National Collaborative Research Infrastructure Strategy. ASKAP uses the resources of the Pawsey Supercomputing Research Centre. Establishment of ASKAP, Inyarrimanha Ilgari Bundara, the CSIRO Murchison Radio-astronomy Observatory and the Pawsey Supercomputing Research Centre are initiatives of the Australian Government, with support from the Government of Western Australia and the Science and Industry Endowment Fund. This paper includes archived data obtained through the CSIRO ASKAP Science Data Archive, CASDA (\url{https://data.csiro.au}).

We used in our work the Astrogeo VLBI FITS image database, DOI: 10.25966/kyy8-yp57, maintained by Leonid Petrov. We acknowledge the use of public data from the Swift data archive. This study uses observations obtained with XMM-Newton, an ESA science mission with instruments and contributions directly funded by ESA Member States and NASA. This paper includes data gathered with the 6.5 meter Magellan Telescopes located at Las Campanas Observatory, Chile. The Pan-STARRS1 Surveys (PS1) have been made possible through contributions of the Institute for Astronomy, the University of Hawaii, the Pan-STARRS Project Office, the Max-Planck Society and its participating institutes, the Max Planck Institute for Astronomy, Heidelberg and the Max Planck Institute for Extraterrestrial Physics, Garching, The Johns Hopkins University, Durham University, the University of Edinburgh, Queen's University Belfast, the Harvard-Smithsonian Center for Astrophysics, the Las Cumbres Observatory Global Telescope Network Incorporated, the National Central University of Taiwan, the Space Telescope Science Institute, the National Aeronautics and Space Administration under Grant No. NNX08AR22G issued through the Planetary Science Division of the NASA Science Mission Directorate, the National Science Foundation under Grant No. AST-1238877, the University of Maryland, and Eotvos Lorand University (ELTE). 

\end{acknowledgments}

\begin{contribution}
Y.C. led the analysis and writing of this work. B.M.G and M.R.D. supervised and oversaw the project progression. D.Z.D. and E.M. contributed to the planning and reduction of the VLA and VLBA observations. T.C., W.P., and E.P. contributed the VLITE/VCSS data that was pivotal to the the discovery of this transient. All other authors (A.D.A. through P.S., ordered alphabetically) contributed to observing, data acquisition, or relevant science expertise.

\end{contribution}

\facilities{VLA, VLBA, ATCA, ASKAP, Swift(XRT), XMM(EPIC), ATT(WiFeS), PS1, Magellan:Baade(IMACS)} 

\software{astropy \citep{2013A&A...558A..33A,2018AJ....156..123A,2022ApJ...935..167A}, CASA \citep[][]{CASA}, Miriad \citep[][]{Sault1995Miriad}, AIPS \citep[][]{AIPS}, HEASoft \citep[][]{HEAsoft}, XSPEC \citep{XSPEC1996}, SAS \citep[][]{SAS}, NumPy \citep[][]{Numpy}, Photutils \citep[][]{Photutils}, SciPy \citep[][]{Scipy}}

\appendix

\section{VLBA Visibilities}\label{apdx:VLBA}

\begin{figure}
    \epsscale{1.0}
    \centering
    \plotone{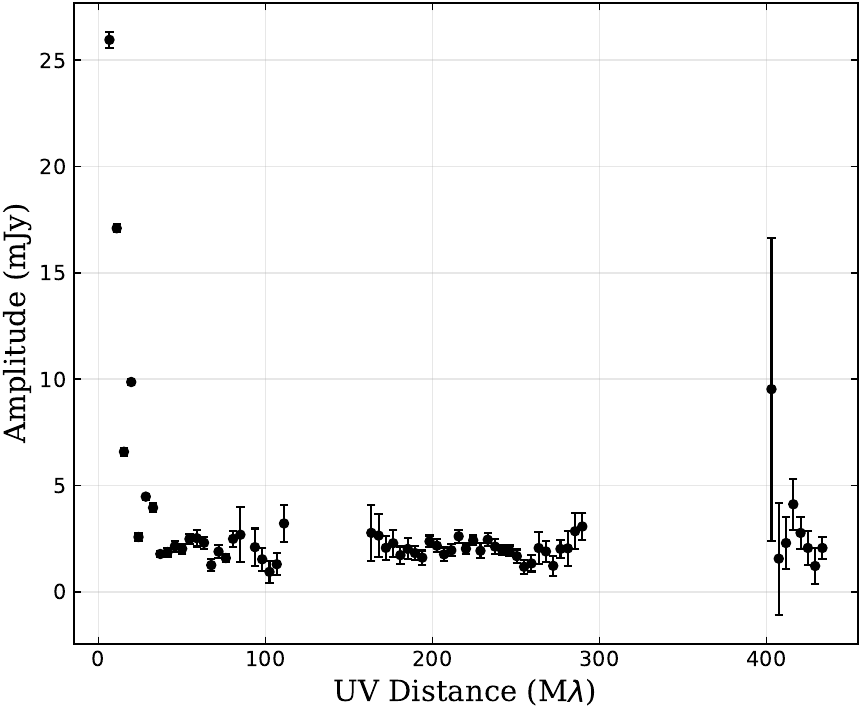}
    \caption{Amplitude over the $(u,v)$ distance of VT\,J1823-2345 from the VLBA Ku-band observation. The visibilities are frequency-averaged, time-averaged, and binned over $(u,v)$ distance for better visualization. The bars reflect variations in the bins.}
    \label{fig:VLBAvis}
\end{figure}

In Figure~\ref{fig:VLBAvis}, we show the calibrated visibilities of VT\,J1823-2345 from the VLBA U-band observations. Structure is clearly seen in the shortest baselines at an amplitude consistent with the flux density measured in the image plane, confirming that VT\,J1823-2345 is resolved. 

\section{Optical Spectra and Archival Photometry of Coincident Star}\label{apdx:OptSpec}

\begin{figure*}
    \centering
    \includegraphics[width=0.88\textwidth]{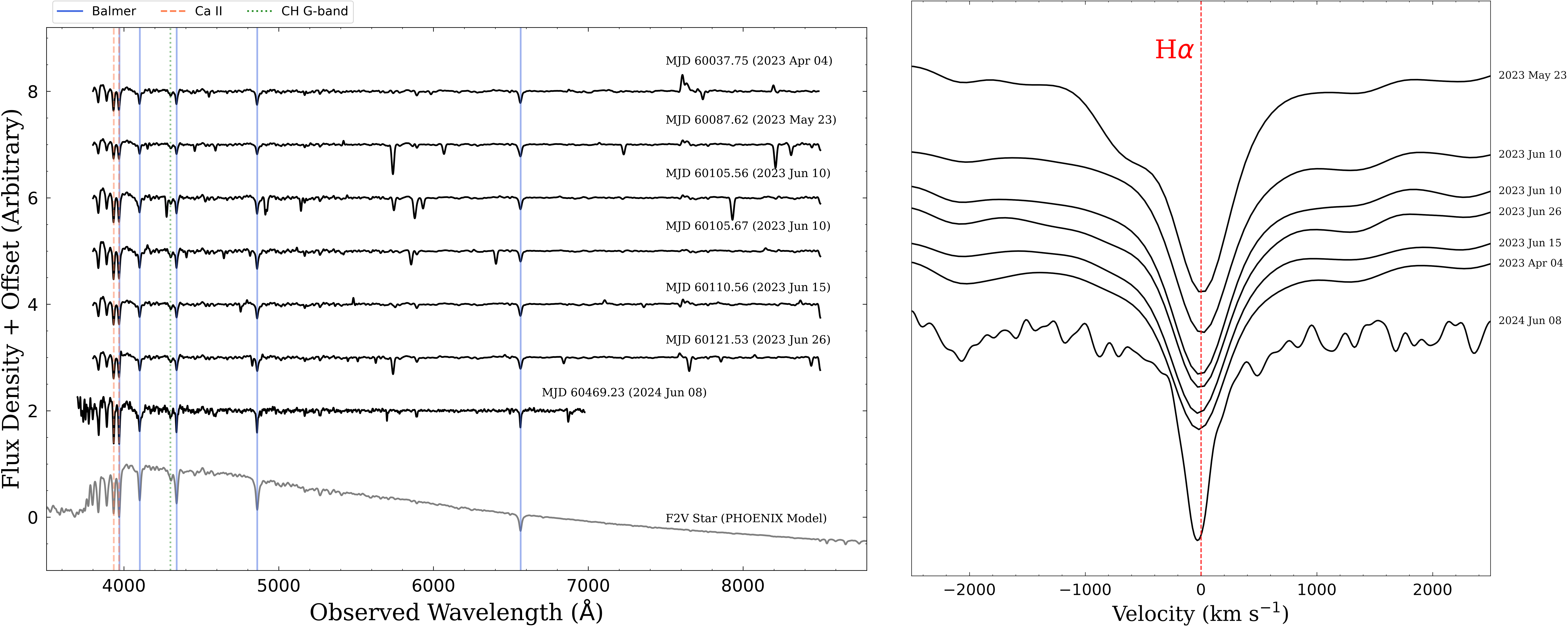}
    \includegraphics[width=0.82\textwidth]{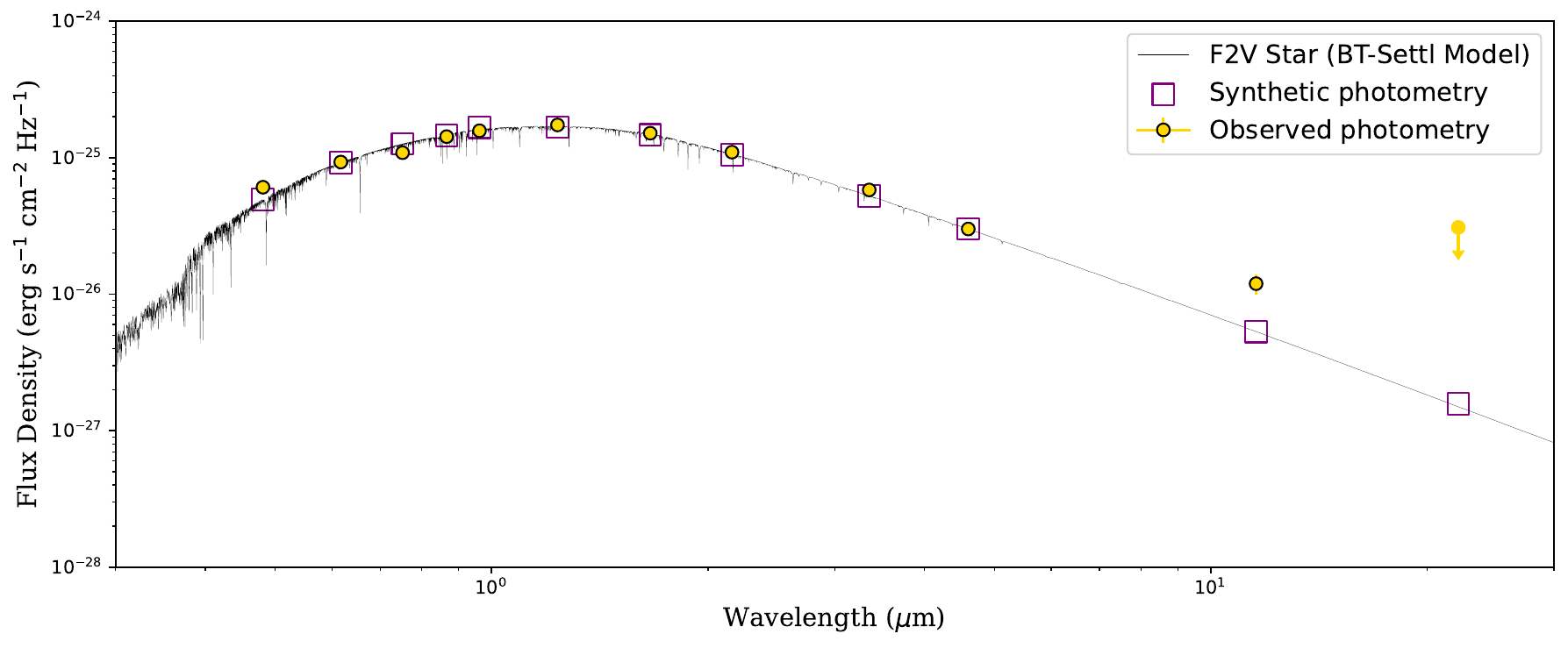}
    \caption{\textit{Top Left}: Optical spectra of the coincident star taken in 2023 and 2024, smoothed for visual clarity. Observation dates are labeled on top of the spectra and vertical lines mark the locations of the Balmer lines, the Ca\,\textsc{II} H \& K lines, and the molecular CH G-band. A PHOENIX model spectrum of an F2V star is also shown at the bottom for comparison. \textit{Top Right}: The same spectra zoomed in to H$\alpha$ (vertical dashed line) shown in velocity space, with observation dates labeled on the right. \textit{Bottom:} Observed PS1, 2MASS, and WISE photometry (circle) compared with the BT-Settl model spectrum of an F2V star and the synthetic photometry at the observed bands (square). Note that the W4 photometry is an upper limit.}
    \label{fig:FstarSpecSED}
\end{figure*}

In the top left panel of Figure~\ref{fig:FstarSpecSED}, we show a series of continuum-subtracted optical spectra of the coincident star taken with WiFeS in 2023 and with IMACS in 2024. In all spectra, we find prominent Balmer absorption lines and Ca\,\textsc{II} H \& K  absorption lines, as expected for F-type stars. We do not find any persistent emission lines that may be expected from an X-ray binary or a background AGN. Note that these spectra show a number of (mainly absorption) lines that appear in one epoch but not in others, which we believe are spurious lines caused by imperfect data reduction (e.g., sky subtraction). 

For comparison, we also show a synthetic spectrum for an F2V star, retrieved from the PHOENIX stellar atmosphere models \citep[][]{Husser2013PHOENIX} with a temperature of $T_{\mathrm{eff}} = 6900\,\mathrm{K}$, surface gravity of $\log(g)=4.5$, and solar metallicity. We selected this spectrum because its absorption features generally match the observed spectra of the coincident star. In particular, this selection is strongly motivated by the ratio between the CH molecular G-band and $\mathrm{H}\gamma$, which is sensitive to the temperature. As seen from this comparison, the observed spectra are broadly consistent with the model spectrum of an F2V star.

In addition, we checked for potential shifts in the spectral lines indicative of binary motion. As an example, we show the series of $\mathrm{H}\alpha$ lines in the top right panel of Figure~\ref{fig:FstarSpecSED}. Based on fitting Gaussian components to the lines, we found variations in the line centroids of up to about $50\,\mathrm{km}\,\mathrm{s}^{-1}$ after barycentric correction. Such variations are broadly consistent with the expected WiFeS wavelength calibration uncertainty of $\sim$\,$50\,\mathrm{km}\,\mathrm{s}^{-1}$ \citep[][]{Childress2014}, and the IMACS spectrum also confirms a relatively low velocity shift. Therefore, we do not have evidence that the F-type star is in a close binary with a hidden companion.

In the bottom panel of Figure~\ref{fig:FstarSpecSED}, we show the PS1 (mean PSF) $grizy$, 2MASS $JHKs$, and WISE $W1W2W3W4$ photometry. For comparison, we also show a BT-Settl \citep[][]{Allard2012BTSettl} model spectrum of an F2V star with $T_{\mathrm{eff}} = 6900\,\mathrm{K}$, $\log(g)=4.5$, and solar metallicity, together with the synthetic photometry at the observed bands. The model spectrum is scaled assuming a radius of $2.5R_\odot$ and a distance of 1.4\,kpc and reddened assuming an extinction of $A_V = 1.64\,\mathrm{mag}$ \citep[][]{SchlaflyFinkbeiner2011}. Note that the scaling is not based on best-fit values, but is chosen to roughly match the observed photometry for comparison. 

The PS1 and 2MASS photometry matches quite well with the model spectrum. Slight deviations in the PS1 photometry are likely due to variations in the images used to construct the mean magnitude. The WISE $W1$ and $W2$ photometry also matches the model spectrum, but excess emission (by a factor of 2) is measured in $W3$. We do not consider this to be strong evidence of a background galaxy because the field is very crowded (Figure~\ref{fig:radoptimg}), so the measured emission in $W3$ may contain IR-bright stars as well as diffuse dust emission at 11.3\,$\mu$m. Therefore, the observed optical-IR SED is overall consistent with the F-type star. 

We also checked the PS1 and WISE photometric light curves taken years prior to the radio discovery. We found no significant fluctuation in the optical or IR light curves at the position of VT\,J1823-2345. Therefore, we have no evidence of a transient flare with an apparent brightness comparable to the F-type star, but we cannot rule out flares hidden below the star. 

We emphasize that the optical and IR data do not rule out background objects such as AGN or a blazar because the stellar continuum could easily hide faint emission features (including emission lines) from distant extragalactic objects. Therefore, the spectra and SEDs presented here are not evidence against an AGN or blazar origin.

\section{Outlines of Alternative, Unlikely Scenarios}\label{sec:unlikelyscene}

In this section, we outline alternative scenarios that we have considered as possible explanations for VT\,J1823-2345, including various Galactic and extragalactic transients as well as blazar variability. First, we consider the expected luminosities, distances, and sizes for the various scenarios. Then, based on these considerations and the observed properties, we present arguments ruling out or disfavoring the alternative scenarios, which ultimately led us to the interpretation that VT\,J1823-2345 is most likely a young PS source.

\subsection{General Considerations on Expected Luminosity, Distance, and Size} \label{subsubsec:LDS}

In the simplest case, if VT\,J1823-2345 is associated with the F-type star (i.e., in a binary), its distance would be $d\sim1.4\,\mathrm{kpc}$ \citep[][]{GaiaDR32023}. Taking the flux density at $\sim$\,GHz frequency to be 70\,mJy, the radio spectral luminosity of VT\,J1823-2345 at 1.4\,kpc would be $L_{\mathrm{GHz}} \sim 10^{13.2}\,\mathrm{W}\,\mathrm{Hz}^{-1}$, which is consistent with jets in X-ray binaries \citep[][]{Pietka2015,Tetarenko2016}. The (major axis) angular size and separation from the VLBA observations would imply a projected emission size of $R_{\mathrm{proj}} \sim 0.6\,\mathrm{AU}$ and a projected separation of $\sim$\,1300\,AU from the F-type star, respectively. While the size could be explained by a compact jet \citep[e.g.,][]{Cao2022AT2019wey}, the large separation argues against accretion in the system (see Appendix~\ref{subsubsec:discussFstar}). Therefore, we also consider alternative scenarios assuming that VT\,J1823-2345 is not associated with the nearby F-type star. 

We can consider the possibility that VT\,J1823-2345 originated from a foreground or background object relative to the F-type star. However, in this case, it is difficult to examine luminosity and size since we lack information on the distance and the exact progenitor or host of the transient. Instead, we derive constraints for a variety of possible scenarios by assuming some typical luminosity range for each scenario, from which we infer a distance and a physical (projected) size. Using the physical size and the time between VLASS epoch 1 and the VLBA observations (1810\,days), we also infer an upper limit on the apparent (average) expansion speed for scenarios involving an expanding shock. Note that for spherical expansions, we assume that the major-axis VLBA size is representative of the shock radius, but we emphasize the caveat that the much smaller minor-axis VLBA size could suggest a physical structure more complex than spherical expansion. For collimated outflows, we assume that the major-axis VLBA size corresponds to the projected linear size of the outflow. We list each scenario in the following text and we compile the calculated values in Table~\ref{tab:LDS}. The calculation uses the ATCA peak flux density $S_\nu \approx 70\,\mathrm{mJy}$ at $\nu_{\mathrm{p,obs}}\sim6.6\,\mathrm{GHz}$, and we note that the derived constraints are order-of-magnitude estimates meant to guide interpretations.

\begin{deluxetable}{cccc}
\tablecolumns{4}
\tablecaption{Assumed Scenarios and Implied Properties}
\tablehead{
Scenario       &
$L_{\mathrm{GHz}}$     & 
$d_\mathrm{L}$ &
$R_{\mathrm{proj}}$ \\
 &
 ($\mathrm{W}\,\mathrm{Hz}^{-1}$) &
 (kpc) &
 (pc) 
}
\startdata
F-type Star & $10^{13.2}$ & 1.4 & $10^{-5.6}$ \\\hline
Stellar Flare & $10^{6}$--$10^{11}$ & $10^{-3.5}$--$10^{-1.0}$ & $10^{-9.2}$--$10^{-6.7}$ \\
Classical Nova & $10^{11}$--$10^{14}$ & $10^{-1.0}$--$10^{0.5}$ & $10^{-6.7}$--$10^{-5.2}$\\
Magnetar Flare & $\sim$\,$10^{15}$ & $\sim$\,$10^{1.0}$ & $\sim$\,$10^{-4.7}$ \\
X-ray Binary & $10^{12}$--$10^{16}$ & $10^{-0.5}-10^{1.5}$ & $10^{-6.2}$--$10^{-4.2}$ \\\hline
SN & $10^{18}$--$10^{22}$ & $10^{2.5}-10^{4.5}$ & $10^{-3.2}$--$10^{-1.2}$  \\
LFBOT & $10^{20}$--$10^{23}$ & $10^{3.5}-10^{5.0}$ & $10^{-2.2}$--$10^{-0.7}$ \\
GRB & $10^{23}$--$10^{25}$ & $10^{5.0}-10^{6.0}$ & $10^{-0.7}$--$10^{0.3}$\\
TDE & $10^{20}$--$10^{25}$ & $10^{3.5}-10^{6.0}$ & $10^{-2.2}$--$10^{0.3}$
\enddata
\tablecomments{See text regarding the scenarios and assumptions.}
\label{tab:LDS}
\end{deluxetable}

Within our Galaxy, a variety of radio transients are known to originate from stellar objects, including active stars and binaries, accreting binaries, and magnetars \citep[e.g.,][]{Mooley2016CNSS}. Note that we do not consider radio transients that only last seconds to minutes. Assuming that VT\,J1823-2345 originated within our Galaxy, we consider the following scenarios:

\underline{\textit{Stellar Flare}:} Active stars and binaries can produce flares at $\sim$\,GHz frequency with $L_{\mathrm{GHz}} \sim 10^{6}-10^{11}\,\mathrm{W}\,\mathrm{Hz}^{-1}$ \citep[][]{GudelBenz1993,BenzGudel1994,Vedantham2022}. If VT\,J1823-2345 is a stellar flare, it should be located at a distance of $d_{\mathrm{L}}\sim10^{-0.5}-10^{2.0}\,\mathrm{pc}$ with an emission size of $R \sim 10^{-3.9}-10^{-1.4}\,\mathrm{AU}$ or $R \sim 10^{-9.2}-10^{-6.7}\,\mathrm{pc}$. At this distance, VT\,J1823-2345 would be a foreground star relative to the coincident F-type star. For flares with lower luminosities, such as M-dwarf flares, the star is expected to be within parsecs.

\underline{\textit{Classical Nova}:} Accreting white dwarfs can produce classical novae with $L_{\mathrm{GHz}} \sim 10^{11}-10^{14}\,\mathrm{W}\,\mathrm{Hz}^{-1}$ \citep[e.g.,][]{Chomiuk2021}, which converts to a distance of $d_{\mathrm{L}}\sim10^{-1.0}-10^{0.5}\,\mathrm{kpc}$ for VT\,J1823-2345. The observed emission size would be $R\sim 10^{-1.4}-10^{0.1}\,\mathrm{AU}$ or $R\sim 10^{-6.7}-10^{-5.2}\,\mathrm{pc}$, which constrains the outflow velocity to be $\langle v \rangle \lesssim 10^{-1.4}-10^{0.1}\,\mathrm{km}\,\mathrm{s}^{-1}$. The inferred size and velocity are unphysical if the radio emission traces a blast wave from a classical nova.

\underline{\textit{Magnetar Giant Flare}:} Magnetars occasionally produce giant flares and their afterglows, based on the well-known case SGR\,1806-20 \citep[][]{Cameron2005,Gaensler2005}, can reach $L_{\mathrm{GHz}} \sim 10^{15}\,\mathrm{W}\,\mathrm{Hz}^{-1}$. If VT\,J1823-2345 is a magnetar giant flare afterglow with a similar radio brightness, it would be located at $d_{\mathrm{L}}\sim10^{1.0}\,\mathrm{kpc}$. The observed emission size would be $R\sim 10^{0.6}\,\mathrm{AU}$ or $R\sim 10^{-4.7}\,\mathrm{pc}$, which constrains the outflow velocity to be $\langle v \rangle \lesssim 10^{0.6}\,\mathrm{km}\,\mathrm{s}^{-1}$. The inferred size and velocity are unphysical if the radio emission traces a blast wave from a magnetar giant flare.

\underline{\textit{X-ray Binary}:} Radio emission from X-ray binaries are associated with jets, and transient behaviors are observed due to jet activities \citep[][]{Done2007}. The observed radio luminosities of X-ray binaries (excluding faint quiescent systems) span roughly a range of $L_{\mathrm{GHz}} \sim 10^{12}-10^{16}\,\mathrm{W}\,\mathrm{Hz}^{-1}$ \citep[e.g.,][]{Tetarenko2016}, which would imply a distance of $d_{\mathrm{L}}\sim10^{-0.5}-10^{1.5}\,\mathrm{kpc}$ for VT\,J1823-2345. At this distance, the projected emission size would be $R_{\mathrm{proj}}\sim 10^{-0.9}-10^{1.1}\,\mathrm{AU}$ or $R_{\mathrm{proj}}\sim 10^{-6.2}-10^{-4.2}\,\mathrm{pc}$. At the most distant and luminous end, the X-ray binary will appear as a microquasar. 

Outside the Galaxy, bright radio emission is observed in energetic explosions and engine-driven outflow, including core-collapse supernovae \citep[CCSNe;][]{Bietenholz2021RadioSN}, luminous fast blue optical transients \citep[LFBOTs;][]{Bright2022}, gamma-ray bursts \citep[GRBs;][]{Chandra2012}, and tidal disruption events \citep[TDEs;][]{Cendes2023}. Assuming VT\,J1823-2345 is an extragalactic transient, we consider the following scenarios:

\underline{\textit{CCSN}:} In core-collapse SNe, bright radio emission is produced by the interaction between the ejecta and the dense surrounding material. The observed radio spectral luminosity spans roughly a range of $L_{\mathrm{GHz}} \sim 10^{18}-10^{22}\,\mathrm{W}\,\mathrm{Hz}^{-1}$ for SNe \citep[][]{Bietenholz2021RadioSN}, implying a distance of $d_{\mathrm{L}}\sim10^{2.5}-10^{4.5}\,\mathrm{kpc}$ for VT\,J1823-2345. At this distance, the projected emission size would be $R\sim 10^{-3.2}-10^{-1.2}\,\mathrm{pc}$, implying an outflow velocity of $\langle v \rangle\lesssim 10^{2.1}-10^{4.1}\,\mathrm{km}\,\mathrm{s}^{-1}$. The inferred outflow velocity at the largest distance ($\sim$\,$10^4\,\mathrm{km}\,\mathrm{s}^{-1}$) is typical of SNe.

\underline{\textit{LFBOT}:} At $\sim$\,GHz frequency, the observed radio spectral luminosity of LFBOTs spans roughly a range of $L_{\mathrm{GHz}} \sim 10^{20}-10^{23}\,\mathrm{W}\,\mathrm{Hz}^{-1}$ \citep[][]{Bright2022}. For VT\,J1823-2345, the implied distance is $d_{\mathrm{L}}\sim10^{3.5}-10^{5.0}\,\mathrm{kpc}$. In this case, the emission size would be $R\sim 10^{-2.2}-10^{-0.7}\,\mathrm{pc}$, which constrains the outflow velocity to be $\langle v \rangle\lesssim 10^{3.1}-10^{4.6}\,\mathrm{km}\,\mathrm{s}^{-1}$. The velocity inferred at the largest distance ($\gtrsim$\,$10^4\,\mathrm{km}\,\mathrm{s}^{-1}$) is consistent with the sub-relativistic outflow of LFBOTs.

\underline{\textit{GRB}:} GRB afterglows produce bright radio emission with $L_{\mathrm{GHz}} \sim 10^{23}-10^{25}\,\mathrm{W}\,\mathrm{Hz}^{-1}$ \citep[][]{Chandra2012}. For VT\,J1823-2345, this luminosity range allows a larger distance of $d_{\mathrm{L}}\sim10^{5.0}-10^{6.0}\,\mathrm{kpc}$ or a redshift range of $z\sim0.02-0.21$. The projected emission size would be $R_{\mathrm{proj}}\sim 10^{-0.7}-10^{0.3}\,\mathrm{pc}$, which implies an average projected outflow speed of $\langle v_{\mathrm{proj}}\rangle \lesssim 10^{4.6}-10^{5.7}\,\mathrm{km}\,\mathrm{s}^{-1}$ or, in terms of the speed of light, $\langle v_{\mathrm{proj}}\rangle \lesssim 0.14c-1.69c$. 

\underline{\textit{TDEs}:} The observed radio spectral luminosities of TDEs span a broad range of $L_{\mathrm{GHz}} \sim 10^{20}-10^{25}\,\mathrm{W}\,\mathrm{Hz}^{-1}$. Within this range, the fainter ones with $L_{\mathrm{GHz}} \sim 10^{20}-10^{23}\,\mathrm{W}\,\mathrm{Hz}^{-1}$ originate from thermal TDEs \citep[][]{Cendes2023}, while the brightest ones with $L_{\mathrm{GHz}} \sim 10^{23}-10^{25}\,\mathrm{W}\,\mathrm{Hz}^{-1}$ are jetted TDEs that launch prompt jets \citep[][]{,Eftekhari2018Sw1644,Rhodes2023AT2022cmc}. Overall, they imply a distance range of $d_{\mathrm{L}}\sim10^{3.5}-10^{6.0}\,\mathrm{kpc}$ for VT\,J1823-2345. The projected emission size would span $R_{\mathrm{proj}}\sim 10^{-2.2}-10^{0.3}\,\mathrm{pc}$ and the average projected outflow speed would be $\langle v_{\mathrm{proj}}\rangle \lesssim 10^{3.1}-10^{5.7}\,\mathrm{km}\,\mathrm{s}^{-1}$.

\subsection{Association with F-type Star and Supernova Remnant} \label{subsubsec:discussFstar}

Our original hypothesis was that VT\,J1823-2345 is associated with the F-type star based on the small angular separation and possibly with the SNR\,G8.7-5.0 since VT\,J1823-2345 is near its projected center. In this case, the radio spectral luminosity of VT\,J1823-2345 would be $L_{\mathrm{GHz}} \sim 10^{13}\,\mathrm{W}\,\mathrm{Hz}^{-1}$, consistent with bright Galactic transients including classical novae and X-ray binaries. This implies that the radio emission is not directly from the star, but from an unseen compact companion. However, our VLBA observations confirm an angular separation of 0.914\,arcsec between VT\,J1823-2345 and the F-type star, posing a significant challenge to our original hypothesis. 

Both classical nova and X-ray binary require accretion onto the compact object, with the F-type star being the likely donor. Since the star is not massive (hence it does not produce significant winds), the accretion should be the result of Roche lobe overflow, and we can use the following formula to approximate the Roche lobe radius \citep[][]{Eggleton1983}:
\begin{equation}
    \frac{r_L}{a} = \frac{0.49q^{2/3}}{0.6q^{2/3} + \ln(1+q^{1/3})},
\end{equation}
where $r_L$ is the Roche lobe radius, $a$ is the orbital separation, and $q$ is the mass ratio of the binary. The observed angular separation would correspond to a (lower limit) orbital separation of $a = 1280\,\mathrm{AU}$. Taking the mass of the F-type star to be $M_* = 1.7\,M_\odot$ \citep[from Gaia DR3;][]{GaiaDR32023}, the Roche lobe radius is $r_L \approx 10^{5}\,\mathrm{R_\odot}$ if the compact object is a white dwarf with $M_{\mathrm{wd}}=1.4\,\mathrm{M_\odot}$ or $r_L \approx 5\times10^{4}\,\mathrm{R_\odot}$ if the compact object is a black hole with $M_\bullet = 20\,\mathrm{M_\odot}$. Note that $r_L$ increases if the mass of the compact object is lower. These values of $r_L$ indicate that it is impossible for the F-type star to transfer mass through Roche lobe overflow over such a wide separation. This is consistent with the lack of accretion disk signatures in the optical spectra (Appendix~\ref{apdx:OptSpec}). Thus, we can confidently rule out accretion from the F-type star as the power source of VT\,J1823-2345.

Originally, we also considered the possibility that the F-type star is associated with SNR\,G8.7-5.0. However, based on the Gaia proper motion, the F-type star is moving in the southwest direction, meaning that it does not trace back to the projected center of the SNR (Figure~\ref{fig:radoptimg}). Although this does not rule out an association, the lack of accretion signatures and the proper motion suggest that the two may not be associated.

\subsection{Galactic Transient}

In Galactic coordinates, VT\,J1823-2345 is located at $l=8.7^{\circ}$ and $b=-4.9^{\circ}$, close to the Galactic plane, and the field around VT\,J1823-2345 is visibly crowded (Figure~\ref{fig:radoptimg}). Therefore, VT\,J1823-2345 could still be associated with a foreground or background Galactic object even if it is not associated with the F-type star. We considered several possible scenarios in Appendix~\ref{subsubsec:LDS} and derived various constraints for Galactic transients, and we now assess whether these are plausible or unlikely scenarios for explaining VT\,J1823-2345.

\underline{\textit{Stellar Flare}:} We note that the brightness temperature of VT\,J1823-2345 ($T_{\mathrm{B}} = 5.7\times10^{9}\,\mathrm{K}$) could allow for flare emission generated by incoherent mechanisms or certain coherent mechanisms with low $T_{\mathrm{B}}$. If VT\,J1823-2345 is a stellar flare, it would realistically be at a distance of $d\lesssim100\,\mathrm{pc}$ (Appendix~\ref{subsubsec:LDS}). At such small distances, one constraint is the lack of bright coincident optical counterpart at the VLBA position. In the $r$-band, we found that the emission of the F-type star can hide a point source with $m_{r}\gtrsim 16\,\mathrm{mag}$ (Section~\ref{subsec:optassoc}), which implies that the optical counterpart of VT\,J1823-2345 should have an absolute magnitude of $M_{r}\gtrsim6\,\mathrm{mag}$ for $d\lesssim100\,\mathrm{pc}$. This brightness is consistent with M dwarfs and K dwarfs \citep[][]{Cifuentes2020}, even if we consider a reddening of $A_r\sim1$--$2\,\mathrm{mag}$ (although such reddening may not be realistic at small distances). This constraint already conflicts with the radio brightness because M-dwarf or K-dwarf flares have relatively low radio luminosities \citep[$L_\nu\lesssim10^8\,\mathrm{W}\,\mathrm{Hz}^{-1}$;][]{Pritchard2024}. In this case, the star should reside within parsecs for $S_\nu = 70\,\mathrm{mJy}$ and appear as an extremely bright foreground star, which is not seen.

Even if we assume that the transient may be a brighter, more distant stellar flare and its progenitor somehow went undetected in the optical, we note that the ratio between the radio and X-ray luminosity of VT\,J1823-2345 is also unusual for stellar flares. Following the G\"udel-Benz relation and assuming $d\lesssim100\,\mathrm{pc}$, we expect the luminosity ratio to be $\log_{10}(L_{\nu,\mathrm{R}}/L_{\mathrm{X}})\lesssim-14$ \citep[][]{BenzGudel1994,Williams2014}, where $L_{\nu,\mathrm{R}}$ is the spectral luminosity at around 5\,GHz, $L_{\mathrm{X}}$ is the soft X-ray luminosity over 0.1--2.4\,keV, and $L_{\nu,\mathrm{R}}/L_{\mathrm{X}}$ should be in units of Hz$^{-1}$. For VT\,J1823-2345, using the ATCA 5\,GHz flux density and the XMM-Newton 0.1--2.4\,keV X-ray flux converted using PIMMS, we find a ratio of $\log_{10}(L_{\nu,\mathrm{R}}/L_{\mathrm{X}})\sim-10.9$. This comparison suggests that the radio luminosity of VT\,J1823-2345 is many orders of magnitude higher than expected and extremely unusual even considering the outliers \citep[][]{Driessen2024}. Another inconsistency lies in the timescale: while stellar flares last $\lesssim$\,day \citep[][]{Mooley2016CNSS,Pietka2015}, VT\,J1823-2345 showed relatively smooth brightening over multiple years. Overall, based on the lack of a bright optical counterpart and the radio and X-ray properties, we argue that VT\,J1823-2345 is highly unlikely to be a stellar flare.

\underline{\textit{Classical Nova}:} If VT\,J1823-2345 is a classical nova, its optical host at larger distances can be faint enough to remain undetected in the PS1 image. In terms of timescale, there exist long-lasting classical novae that evolve over $\gtrsim$\,10$^{3}$\,days \citep[][]{Chomiuk2021}, but VT\,J1823-2345 with $t_{1/2}\sim10\,\mathrm{yr}$ would represent a very extreme case if it is a classical nova. The brightness temperature of VT\,J1823-2345 of $T_{\mathrm{B}} \sim 10^{9}\,\mathrm{K}$ is also orders of magnitude higher than $T_{\mathrm{B}} \lesssim 10^2-10^{7}\,\mathrm{K}$ observed in classical novae \citep[][]{Chomiuk2021}. Perhaps the most stringent constraint is the inferred size that leads to an unphysical $\langle v \rangle \lesssim 1\,\mathrm{km}\,\mathrm{s}^{-1}$ for VT\,J1823-2345 considering the scenario of a classical nova (Appendix~\ref{subsubsec:LDS}). Therefore, we strongly disfavor classical nova as a plausible explanation for VT\,J1823-2345.

\underline{\textit{Magnetar Giant Flare}:} In the well-known case of SGR\,1806-20 \citep[][]{Gaensler2005}, the afterglow of the magnetar giant flare evolved over weeks and showed an expanding nebula with an angular size of tens of mas (at $d\sim10\,\mathrm{kpc}$). These properties are entirely inconsistent with VT\,J1823-2345, which evolved over many years and maintained a compact size of 0.4\,mas. These also lead to an unphysical $\langle v \rangle \lesssim 10\,\mathrm{km}\,\mathrm{s}^{-1}$ for VT\,J1823-2345 (Appendix~\ref{subsubsec:LDS}). Although variations are naturally expected in the flares of different magnetars, it is unlikely that variations on the population level can explain such extreme differences (which would require extreme energetics and environment). Therefore, based on these discrepancies, we strongly disfavor magnetar giant flare as a plausible explanation for VT\,J1823-2345.

\underline{\textit{X-ray Binary}:} The radio luminosities of active X-ray binaries and microquasars allow a distance up to tens of kpc for VT\,J1823-2345 (Appendix~\ref{subsubsec:LDS}), where the optical counterpart could be too faint to detect. X-ray binaries also appear as luminous radio transients during state transitions and outbursts \citep[][]{Done2007}. However, there are a number of discrepancies between the properties of VT\,J1823-2345 and those typically observed for X-ray binaries. First and foremost, the outburst durations from X-ray binaries are typically weeks to years \citep[][]{Tetarenko2016WATCHDOG}, while the time above half-maximum of VT\,J1823-2345 is already a decade. Second, the observed luminosity ratio of VT\,J1823-2345 at distances reasonable for X-ray binaries is $\nu L_{\nu,\mathrm{R}}/L_{\mathrm{X}}\sim0.05$, which is about 3--7 orders of magnitude higher than the known radio and X-ray correlation of Galactic black hole X-ray binaries \citep[][]{Gallo2003,Gallo2006}, although outliers may exist \citep[e.g.,][]{Tetarenko2016}. This discrepancy is worse in the case of neutron star X-ray binaries, which are generally fainter in the radio \citep[][]{Migliari2006}. Third, the X-ray flux of VT\,J1823-2345 did not change during the radio brightening, which is again inconsistent with the expected radio and X-ray coupling in X-ray binaries. We note that the low X-ray flux is unlikely due to absorption because we found a moderately hard X-ray photon index ($\Gamma_\mathrm{X}\sim1.5$) whereas significant suppression of soft X-ray is expected for extreme amounts of absorption by neutral hydrogen. 

In addition, we note that X-ray binaries typically show flat or inverted spectra at $\sim$\,GHz frequencies, interpreted as the superposition of self-absorbed components from different portions of a continuous jet \citep[][]{Blandford&Konigl1979}. The persistently peaked spectrum of VT\,J1823-2345 is perhaps more consistent with a transient jet (a discrete ejection) showing a single self-absorbed synchrotron component. However, this interpretation appears to conflict with the compact VLBA size, which implies an apparent average velocity of $\langle v_{\mathrm{proj}} \rangle \lesssim 16\,\mathrm{km}\,\mathrm{s^{-1}}$ for the ejected material (Appendix~\ref{subsubsec:LDS}). 

Overall, we find a number of discrepancies between VT\,J1823-2345 and typical X-ray binaries. We note that there may exist extreme or fine-tuned scenarios that may explain VT\,J1823-2345 as a Galactic black hole. For example, if VT\,J1823-2345 is a transient jet, an inclination angle of almost exactly 0$^{\circ}$ (face-on jet) may explain the compact size. According to the fundamental plane of black hole activity \citep[][]{Merloni2003,Gultekin2019}, the radio to X-ray luminosity ratio of VT\,J1823-2345 is consistent with an intermediate-mass black hole with a mass of $\sim$\,$10^5\,M_\odot$ at $\sim$\,10\,kpc. However, rather than adopting exotic or fine-tuned explanations, we argue based on the overall discrepancies discussed above that VT\,J1823-2345 is unlikely associated with a Galactic X-ray binary or black hole.

\subsection{Extragalactic Non-relativistic Transient}

If VT\,J1823-2345 is not a Galactic transient, then it must have originated from outside the Galaxy. Various classes of extragalactic transients are known to produce bright radio synchrotron emission from shock interaction. Here, based on the constraints derived in Appendix~\ref{subsubsec:LDS}, we consider the non-relativistic transients: CCSNe, LFBOTs, and thermal TDEs.

\underline{\textit{CCSN or LFBOT}:} Based on the assumed radio luminosities, we inferred a distance for VT\,J1823-2345 of $d_{\mathrm{L}}\lesssim30\,\mathrm{Mpc}$ in the case of a CCSN and $d_{\mathrm{L}}\lesssim100\,\mathrm{Mpc}$ in the case of an LFBOT. Realistically, VT\,J1823-2345 is likely on the most luminous and distant end such that the inferred outflow velocity is reasonable for a SN ($\langle v \rangle \sim 10^{4}\,\mathrm{km}\,\mathrm{s^{-1}}$) or for an LFBOT ($\langle v \rangle \gtrsim 0.1c$). We note that distances of tens of Mpc are extremely nearby for SNe and LFBOTs. Based on the volumetric rates, we expect $\sim$\,11 CCSNe per year within 30\,Mpc \citep[][]{Perley2020}, although only a small fraction of CCSNe are radio-bright \citep[][]{Bietenholz2021RadioSN,Sfaradi2025SN}, and $\sim$\,0.001--1.7 LFBOT is expected per year within 100\,Mpc \citep[although only one LFBOT is known to be within 100\,Mpc so far;][]{Ho2023LFBOT}. 

Despite the small distance, no optical flare is seen in the ATLAS light curve between 2015 and 2025 or in PS1 between 2009 and 2014. The absolute $r$-band magnitude is expected to be roughly between $-16$ and $-20$ for SNe \citep[][]{Perley2020} and between $-19$ and $-21$ for LFBOTs \citep[][]{Margutti2019,Perley2026}, which translates to an extremely bright $r$-band apparent magnitude between 16.4 and 12.4 for a SN at 30\,Mpc and between 16.0 and 14.0 for an LFBOT at 100\,Mpc. Furthermore, we note that the optical image does not show any extended emission from a potential host galaxy beyond $\sim$\,1--2\,arcsec of VT\,J1823-2345. We note that at the position of VT\,J1823-2345, foreground Galactic extinction \citep[][]{SchlaflyFinkbeiner2011} is $A_\mathrm{r}\sim1.4$, which still allows a SN or an LFBOT to be detected at 17th to 14th magnitude in r-band. In Section~\ref{subsec:optassoc}, we found that the foreground emission of the F-type star could hide a galaxy with an integrated magnitude of $m_r \gtrsim 15$ and a half-light radius of $R_e\sim1.25\,\mathrm{arcsec}$, which corresponds to $R_e \lesssim600\,\mathrm{pc}$ at $d_\mathrm{L} \lesssim 100\,\mathrm{Mpc}$, consistent with a dwarf galaxy \citep[][]{McConnachie2012}. These findings imply that if VT\,J1823-2345 is a SN or an LFBOT, it must be an extreme nearby case with no optical counterpart (perhaps missed or obscured) and no visible host galaxy (perhaps faint, compact or obscured), which is possible but the allowed scenarios are limited.

There are also some possible discrepancies between the properties of VT\,J1823-2345 and those typically observed in CCSNe and LFBOTs. The time above half-maximum of a decade of VT\,J1823-2345 is much longer than the typical duration of months to years observed in the radio light curves of SNe and LFBOTs \citep[][]{Bietenholz2021RadioSN,Bright2022}. We can effectively rule out the case of a typical LFBOT, because all known LFBOTs faded within a few hundred days with the peak frequency falling below $\sim$\,GHz frequency \citep[][]{Ho2019,Bright2022,Nayana2025,Sevilla2026}, which is inconsistent with VT\,J1823-2345. Some Type II SNe (mainly Type IIn SNe) are known to rise over thousands of days and evolve over decades \citep[from interaction with dense, extended circumstellar material;][]{Bietenholz2021RadioSN}, although such SNe constitute only $\sim$\,10\% of all CCSNe \citep[][]{Perley2020}. However, we note that these long-lasting SNe typically become optically thin with peak frequencies below GHz frequency within a few thousand days \citep[e.g.,][]{Ryder1993,Bartel2002,Meunier2013,Chandra2020}, in contrast to the $\sim$\,6\,GHz peak of VT\,J1823-2345 after at least $\sim$\,2000\,days. 

We can also examine this from a physical perspective. Using standard synchrotron modeling for SNe \citep[following Equation~(23) of][assuming equipartition, $f=0.5$, and the microphysics parameter $\epsilon_B=1/3$]{Ho2019}, we find a high progenitor mass-loss rate of $\dot{M}\sim(2-5)\times10^{-4}\,M_\odot\,\mathrm{yr}^{-1}$ from the radio spectrum of VT\,J1823-2345 assuming a wind speed of $v_w=10\,\mathrm{km}\,\mathrm{s}^{-1}$ (appropriate for red supergiants) and a post-explosion time of 2000--3000\,days. Note that the mass-loss rate can be orders of magnitude larger if the wind speed is higher (e.g., in the case of Wolf-Rayet stars). Taking the VLBA size to be the shock radius and assuming free expansion (implying $v_{\mathrm{SN}}\sim7000$--$11000\,\mathrm{km}\,\mathrm{s}^{-1}$ at 30\,Mpc) and constant mass loss, we find the mass of the swept-up circumstellar material to be $M_{sw}\sim1-3\,M_\odot$ and the kinetic energy in the swept-up mass to be $E_{sw}\sim1.7\times10^{51}\,\mathrm{erg}$. This finding is potentially unphysical because $E_{sw}$ is already on par with the total kinetic energy of a SN ($10^{51}\,\mathrm{erg}$) even though the radio SEDs were taken while VT\,J1823-2345 was still evolving. Although we cannot rule out that certain extreme parameters/conditions may still reasonably describe VT\,J1823-2345, our findings overall suggest that VT\,J1823-2345 must be an exceptionally extreme (possibly unphysical) case if it is a SN or an LFBOT. Rather than adopting this interpretation, we regard these findings as evidence that VT\,J1823-2345 is highly unlikely to be a SN or an LFBOT.

\underline{\textit{Thermal TDE}:} Most known TDEs are thermal TDEs (as opposed to jetted TDEs), and while few exhibit prompt radio emission \citep[][]{Alexander2020}, a large fraction of thermal TDEs are now known to show late-time radio brightening \citep[][]{Sfaradi2022,Cendes2023,Alexander2025}. The observed radio spectral luminosities of thermal TDEs are similar to those of LFBOTs (at $\sim$\,GHz frequency), which implies a similar distance constraint of $d_\mathrm{L} \lesssim 100\,\mathrm{Mpc}$. Assuming a volumetric rate of $\sim$\,$3\times10^{-7}\,\mathrm{Mpc}^{-3}\,\mathrm{yr}^{-1}$ \citep[from optically-selected TDEs;][]{Yao2023}, we expect about 1.3 TDEs per year within 100\,Mpc, although we note that the rate is likely smaller if we assume that the host is a dwarf galaxy from the optical non-detection. The lack of an observed optical flare for VT\,J1823-2345 is less problematic in the case of a thermal TDE if the radio brightening occurred years or decades after the event.

Empirically, the radio emission of thermal TDEs is known to evolve rapidly and their peak frequencies fall to or below $\sim$\,GHz within a timescale of months or years after the brightening \citep[e.g.,][]{Christy2024,Christy2025,Burn2025,Goodwin2025,Goodwin2025XrayTDE,Hajela2025,Sfaradi2025}, although there are some cases with longer timescales \citep[e.g.,][]{Golay2025,Cendes2026AT2018hyz}. In general, the slow temporal and spectral evolution, particularly the persistent $\sim$6\,GHz peak, would make VT\,J1823-2345 an outlier amongst thermal TDEs. The exact origin of the radio emission in thermal TDEs is still debated, and potential explanations include accretion state transition \citep[][]{Goodwin2026}, disk instabilities \citep[][]{Piro&Mockler2025}, structures in the surrounding gas \citep[][]{MatsumotoPiran2024,Zhuang2025}, and off-axis or choked jets \citep[][]{Matsumoto2023,Teboul2023,Lu2024}. If VT\,J1823-2345 is a thermal TDE, it likely involves a non-relativistic outflow given the constraint on the average apparent expansion speed $\langle v \rangle \lesssim 0.16c$. 

We do not attempt to examine all possible scenarios, but we can examine physical parameters with the same synchrotron modeling. Based on the 6\,GHz peak and assuming a distance of 100\,Mpc and a time of 2000--3000\,days, we find an ambient density of $n_{\mathrm{ext}} \sim (2-4)\times10^4\,\mathrm{cm}^{-3}$ \citep[following Equation~(16) of][assuming equipartition, $f=0.5$, and $\epsilon_B=1/3$]{Ho2019}. This density (at a radius equal to the VLBA size) is orders of magnitude higher than those inferred from thermal TDEs \citep[][]{Cendes2023}, which is a result of the high peak frequency and long timescale. It also implies a swept-up mass of $M_{sw}\sim30-130\,M_\odot$ assuming a density profile of $\rho\propto R^{k}$ where $k$ is between $-1$ and $-2$. This is likely unphysical since the outflow mass should be a (small) fraction of the disrupted stellar mass and with the swept-up mass, it would require an enormous amount of kinetic energy to maintain an expansion speed of $\gtrsim10^4\,\mathrm{km}\,\mathrm{s}^{-1}$. We note that a smaller distance can lower $M_{sw}$ (although $n_{\mathrm{ext}}$ would increase by more than an order of magnitude), but it would again imply an exceptionally extreme case at a very small distance for an intrinsically rare transient. Similar to the case of SNe and LFBOTs, we consider these findings to be evidence that VT\,J1823-2345 is highly unlikely to be a thermal TDE.

\subsection{Gamma-Ray Burst}

From the radio luminosities of GRB afterglows, we found a larger distance range of $d_\mathrm{L} \lesssim 1\,\mathrm{Gpc}$ ($z\lesssim0.2$) for VT\,J1823-2345 compared to the cases for SNe and LFBOTs. At 1\,Gpc, a galaxy with an effective radius of a few kpc would appear compact (smaller than an arcsec) in the PS1 image and could be hidden if $M_r \gtrsim -23\,\mathrm{mag}$. Therefore, in this case, a range of galaxies \citep[][]{Shen2003} may be undetected at the location of VT\,J1823-2345 in the PS1 image, so the lack of a visible host may not be an issue.

For the GRB interpretation, the main discrepancy lies in the radio properties. Most long-duration GRBs evolve over a timescale of hundreds of days and their light curves peak over weeks to months at $\sim$\,GHz frequencies \citep[][]{Chandra2012}. There are much fewer detections of radio afterglows of short-duration GRBs, but these are generally fainter and faster than long-duration GRBs \citep[][]{Fong2015}. In the standard afterglow model, the peak frequency is expected to decrease over time \citep[][]{GranotSari2002}, and for typical long-duration GRBs, both the characteristic frequency $\nu_m$ and self-absorption frequency $\nu_{a}$ are expected to be well below 1\,GHz at $t\gtrsim10^3\,\mathrm{days}$ \citep[][]{Leung2023,Chen2024InvertedTransient}. 

Extreme absorption is likely required to explain the high-frequency peak of VT\,J1823-2345. If we assume $\nu_a/(1+z)=6\,\mathrm{GHz}$, $t=2000\,\mathrm{days}$, $z=0.2$, and typical parameters for long-duration GRBs \citep[following Section~4.1.2 of][]{Chen2024InvertedTransient}, then we find an extreme ambient density of $n\sim4\times10^5\,\mathrm{cm}^{-3}$ assuming a uniform density profile \citep[from spectrum 2 of][]{GranotSari2002}. This is orders of magnitude larger than the interstellar medium density ($n_{\mathrm{ISM}}\sim10^{-1}-10^1\,\mathrm{cm^{-3}}$) typically seen in long-duration GRBs \citep[][]{Chandra2012}. This density also implies a swept-up mass of $M_{sw}\sim10^5f_{\theta}\,M_\odot$ within the observed VLBA size, where $f_\theta = (\theta_j\Gamma)^2$ is the filling factor, $\theta_j$ is the jet opening angle, and $\Gamma$ is the Lorentz factor \citep[][]{BarniolDuran2013}. Unless the jet is extremely narrow (with extremely small $\theta_j$) over many years, this swept-up mass is unphysical because the total kinetic energy of the jet ($10^{51}\,\mathrm{erg}$) is insufficient for accelerating this material and producing the bright radio afterglow. We acknowledge that we have not considered all possible parameters (e.g., microphysics) or exotic scenarios (e.g., a GRB involving prolonged energy injection), which can add significant complexity to the modeling. However, we argue based on the extraordinary difference between VT\,J1823-2345 and known GRBs that VT\,J1823-2345 is highly unlikely to be a GRB afterglow.

\subsection{Blazar Variability}\label{subsec:blazarvar}

Blazars are known to show dramatic variability at centimeter wavelengths over a timescale of months and years \citep[e.g.,][]{Aller2017,Angelakis2019}, commonly attributed to shocks within the jet or changes in the jet orientation \citep[see review by][]{Raiteri2025}. They can also exhibit peaked radio spectra during flares and are often regarded as contaminants in reported samples of PS sources, particularly for HFP sources \citep[][]{Tinti2005,Torniainen2005,Orienti2007,Orienti2008}. These characteristics are similar to the observed radio properties of VT\,J1823-2345. 

It is difficult to definitively rule out blazar variability for VT\,J1823-2345 since we lack information on the optical host and the exact radio morphology. However, one inconsistency is perhaps seen in the brightness temperature. A defining characteristic of blazars is their relativistically beamed emission, and the resulting Doppler boost can significantly increase the observed brightness temperature. Based on the VLBA observations, we measured $T_{\mathrm{B}}/(1+z) = 5.7\times10^{9}\,\mathrm{K}$ for VT\,J1823-2345, which yields an order-of-magnitude estimate of $T_\mathrm{B} \sim 10^{10}\,\mathrm{K}$ at cosmological distances ($z\gtrsim1.0$). The brightness temperature measured for blazars, however, is typically $T_{\mathrm{B}}\sim10^{11}-10^{13}\,\mathrm{K}$, with a very small fraction of blazars showing $T_{\mathrm{B}} \lesssim10^{10}\,\mathrm{K}$ \citep[][]{Kovalev2005,Homan2021}. Assuming an intrinsic brightness temperature of $T_{\mathrm{B,int}}=4.1\times10^{10}\,\mathrm{K}$ \citep[][]{Homan2021}, the typical Doppler factor for blazars is $\delta = T_{\mathrm{B}}/T_{\mathrm{B,int}}\sim1-100$. For VT\,J1823-2345, the Doppler factor $\delta = 0.14(1+z)$ is small and likely implies deboosting, in which case the angle to the line of sight may be large ($\theta \gtrsim20^{\circ}$) for typical Lorentz factors \citep[$\Gamma \lesssim50$; see Equation~5 and 6 of][]{Homan2021}. This interpretation may indicate that beaming is weak in VT\,J1823-2345 and may argue against the blazar interpretation. 

In the case of HFP sources, radio polarization has been used to argue in favor of a blazar origin \citep[e.g.,][]{Orienti2008}. Specifically, while genuine PS sources tend to be unpolarized or weakly polarized \citep[likely due to Faraday depolarization as expected of compact jets;][]{ODeaSaikia2021}, significant polarization (with linear polarization fractions $\gtrsim$\,1\%) has been observed for a large fraction of HFPs that show significant variability and core-jet or unresolved radio morphology, likely associated with blazars \citep[][]{Orienti2008}. This is consistent with the bulk blazar population, which shows a polarization fraction of $\sim$\,1--3\,\% at centimeter wavelengths \citep[e.g.,][]{Linford2012,Hodge2018}, although a small fraction of blazars does show lower linear polarization fractions. In comparison, VT\,J1823-2345 is weakly polarized, with a linear polarization fraction of $p_{\mathrm{L}}\simeq0.73\%$ at 8.8\,GHz and $p_{\mathrm{L}}<0.25\%$ at 5.5\,GHz, which is more consistent with PS sources. The decrease in polarization fraction over wavelength also implies intrinsic depolarization as expected of genuine PS sources. Furthermore, we note that most blazars show $|\mathrm{RM}|\lesssim1000\,\mathrm{rad}\,\mathrm{m}^{-2}$ in the observer frame \citep[e.g.,][]{Hovatta2012}. For VT\,J1823-2345, we found an extreme $\mathrm{RM_{obs}}\sim12,000\,\mathrm{rad}\,\mathrm{m}^{-2}$ in the observer frame, which is perhaps more consistent with some PS sources that do show extreme values of $|\mathrm{RM}|\gtrsim1000-10000\,\mathrm{rad}\,\mathrm{m}^{-2}$ \citep[][]{ODea1998,ODeaSaikia2021}, although even as a PS source, VT\,J1823-2345 would represent an extreme case in terms of polarization properties. Overall, the polarization properties of VT\,J1823-2345 perhaps also argue against the blazar interpretation.

In some cases, studies have also argued in favor of blazar variability to explain rapid changes in the radio spectrum. In particular, rapid changes in spectral shape (i.e., transitions between flat and peaked spectra) are likely a strong indicator of blazar variability \citep[e.g.,][]{Tinti2005,Torniainen2005,Orienti2007,Ross2021}. In the case of VT\,J1823-2345, we did not find significant changes in the spectral shape over a timescale of about a year (Figure~\ref{fig:RadioSEDs}), although we cannot rule out changes over a longer timescale. Even when the spectral shape remains peaked over time, rapid evolution in the spectral peak is often considered to be inconsistent with an expanding jet \citep[e.g.,][]{Tinti2005,Orienti2007}. As an illustration, we follow the same analysis in Section~5 of \citet{Tinti2005} for identifying blazar contamination in PS sources by considering the expected correlation between the fractional change in peak frequency and the fractional change in age, $\Delta\nu_{\mathrm{p}}/\nu_{\mathrm{p}} \propto \Delta t/t$, for a young, expanding jet. Although we did not measure a statistically significant change in $\nu_\mathrm{p}$ for VT\,J1823-2345, if we assume that it decrease by 0.5\,GHz, then we find $\Delta\nu_{\mathrm{p}}/\nu_{\mathrm{p}}\sim-0.08$ between 2023 and 2024, which corresponds to $\Delta t/t \sim0.07$. In the observer frame, taking $\Delta t=1.3\,\mathrm{yr}$, we find $t\sim20\,\mathrm{yr}$ for the age of VT\,J1823-2345. Although this would be considered unrealistic for many PS sources with ages $\gtrsim$\,10$^{2}$\,--\,10$^{4}$\,yr \citep[as argued by][]{Tinti2005}, it is broadly consistent with the interpretation that VT\,J1823-2345 is a transient evolving over decades based on the brightening between NVSS and VLASS. Note that if the change in $\nu_{\mathrm{p}}$ is smaller than 0.5\,GHz, then the inferred age can be longer than 20\,yr.

Therefore, we do not find any strong motivations to adopt an a priori assumption that VT\,J1823-2345 is associated with blazar variability, especially if VT\,J1823-2345 can be reasonably explained as a young transient. Based on the arguments above, we disfavor blazar variability as the interpretation for VT\,J1823-2345, but we recognize that the available data does not fully rule out a blazar origin. Additional monitoring of the radio continuum (e.g., for changes in spectral shape) and VLBI observations will be crucial for ruling out the blazar interpretation.

\bibliography{ref}{}
\bibliographystyle{aasjournalv7}

\end{document}